\documentclass[useAMS,usenatbib]{mn2e}

\usepackage{graphicx}
\usepackage{amsmath}
\usepackage{rotating,booktabs}
\usepackage{graphicx}
\usepackage[T1]{fontenc}
\usepackage[utf8]{inputenc}
\usepackage{mathtools}  
\usepackage{amssymb}
\usepackage{tabulary}
\usepackage{booktabs}
\usepackage{subfigure}
\usepackage{minitoc}
\usepackage[T1]{fontenc}
\usepackage{graphicx}
\usepackage{draftcopy}
\usepackage{algorithm}
\usepackage{algpseudocode}
\usepackage{pifont}
\usepackage{relsize}
\usepackage[T1]{fontenc}
\usepackage[utf8]{inputenc}
\usepackage[usenames,dvipsnames,svgnames,table]{xcolor}
\newcommand{\ee}{\mathrm{e}}
\newcommand{\ii}{\mathrm{i}}
\usepackage[usenames,dvipsnames,svgnames,table]{xcolor}
\usepackage{enumerate}

\newcommand{\ATM}[1]{\textcolor{black}{{#1}}}
\newcommand{\ATMX}[1]{\textcolor{black}{{#1}}}
\newcommand{\ATMR}[1]{\textcolor{black}{{#1}}}
\newcommand\norm[1]{\left\lVert#1\right\rVert}
\newcommand{\Rd}{\textup{\textrm{d}}}
\usepackage{mwe}    
\usepackage{subfig}

\usepackage[usenames,dvipsnames,svgnames,table]{xcolor}

\newcommand{\Refcom}[1]{\textcolor{black}{{#1}}}
\newcommand{\Refnew}[1]{\textcolor{black}{{#1}}}
\newcommand{\RefEq}[1]{\textcolor{black}{{#1}}}

\title[Baseline-dependent sampling and BDWFs]{Baseline-dependent \Refcom{sampling and windowing} for radio interferometry: data compression, field-of-interest shaping and outer field suppression}
\author[M. Atemkeng et \textit{al.}]
{M. Atemkeng$^{1}$\thanks{E-mail: m.atemkeng@gmail.com }, O. Smirnov$^{1,2}$, C. Tasse$^{1,3}$, G. Foster$^{4,5}$, A. Keimpema$^{6}$, \\\\
{\hspace{-0.09cm}\LARGE$\mathrm{Z.~Paragi^{6}~and~J.~Jonas^{1,2}}$}\\
$^1$Department of Physics and Electronics, Rhodes University, PO Box 94, Grahamstown, 6140, South Africa\\
$^2$SKA South Africa, 3rd Floor, The Park, Park Road, Pinelands, 7405, South Africa\\
$^3$GEPI, Observatoire de Paris, CNRS, Universite Paris Diderot, 5 place Jules Janssen, 92190 Meudon, France\\
$^4$University of Oxford, Sub-Department of Astrophysics, Denys Wilkinson Building, Keble Road, Oxford, OX1 3RH, UK\\
$^5$Department of Astronomy, University of California, Berkeley, 501 Campbell Hall \#3411, Berkeley, CA, 94720, USA\\
$^6$Joint Institute for VLBI ERIC in Europe, Oude Hoogeveensedijk 4, 7991 PD Dwingeloo, Netherlands
}
\begin{document}

\date{Accepted 2018 March 02. Received 2018 February 07; in original form 2017 October 07}

\pagerange{\pageref{firstpage}--\pageref{lastpage}} \pubyear{2018}

\maketitle

\label{firstpage}

\begin{abstract}
Traditional radio interferometric correlators produce regular-gridded samples of the true $uv$-distribution by averaging the signal over constant, discrete time-frequency intervals.
This  \Refcom{regular sampling} and averaging then translate to be irregular-gridded samples in the $uv$-space, and results in a baseline-length-dependent loss of amplitude and phase coherence, which is dependent on the distance from the image phase centre. 
The effect is often referred to as ``decorrelation'' in the $uv$-space, which is equivalent in the source domain to ``smearing''.
\Refcom{This work discusses and implements a regular-gridded
sampling scheme in the $uv$-space (baseline-dependent sampling) and  windowing} that allow for data compression, field-of-interest shaping and source suppression. \Refcom{The baseline-dependent sampling}
requires irregular-gridded sampling in the time-frequency \Refnew{space} i.e. \Refnew{the time-frequency interval becomes baseline-dependent. Analytic models and simulations are used to show that decorrelation remains constant across all the baselines when applying  baseline-dependent sampling and windowing.} 
Simulations using MeerKAT telescope and the  European Very Long Baseline Interferometry Network  show that both data compression, field-of-interest shaping and outer field-of-interest suppression \Refcom{are} achieved. 
\end{abstract}
\newcommand{\VV}{\mathcal{V}}
\newcommand{\PP}{\mathcal{P}}
\newcommand{\VVM}{\textcolor{black}{\widetilde{V}}}
\newcommand{\VVMF}{\textcolor{black}{\widetilde{\mathcal{V}}}}

\newcommand{\WW}{\mathcal{W}}
\newcommand{\II}{\mathcal{I}}
\newcommand{\IID}{\mathcal{I}^\mathrm{D}}
\newcommand{\IIA}{\mathcal{I}^\mathrm{A}}
\newcommand{\IIDI}{\mathcal{I}^\mathrm{DI}}
\newcommand{\EE}{\mathcal{E}}
\newcommand{\FF}{\mathcal{F}}
\newcommand{\HH}{\mathcal{H}}
\newcommand{\TT}{\mathcal{T}}
\newcommand{\NN}{\mathcal{N}}
\newcommand{\uu}{\bmath{u}}
\newcommand{\Btf}{\mathsf{B}^{[\Delta t\Delta\nu]}}
\newcommand{\Babtf}{\mathsf{B}^{[\alpha\Delta_{pq} t,\beta\Delta_{pq}\nu]}}
\newcommand{\Bab}{\mathsf{B}^{[\alpha\beta]}}
\newcommand{\Buv}{\mathsf{B}^{[uv]}}
\newcommand{\Bij}{\mathsf{B}}
\newcommand{\Ptf}{\Pi^{[t\nu]}}
\newcommand{\Puv}{\Pi^{[uv]}}
\newcommand{\Vm}{V^\mathrm{M}}
\newcommand{\Vs}{V^\mathrm{S}}
\newcommand{\BIN}[2]{#1 s $\times$ #2 MHz}
\newcommand{\Bda}{\mathsf{B}^{[\Delta_{\uu_{pq}} t, \Delta_{\uu_{pq}} \nu]}}
\newcommand{\Dbda}{\mathsf{D}^{[\Delta_{pq} t, \Delta_{pq} \nu]}}
\newcommand{\Dbdaph}{\mathsf{D}^{[\Delta_{\alpha \beta} t, \Delta_{\alpha \beta} \nu]}}
\newcommand{\WF}[3]{{#1}$#2${}$\times${}$#3$} 
\newcommand{\CF}[2]{CF=#1$\times$#2}

\newcommand{\JVLA}{\textcolor{black}{\text{VLA}}}
\newcommand{\Bdaph}{\mathsf{B}^{[\Delta_{\uu_{\alpha \beta}} t, \Delta_{\uu_{\alpha \beta}} \nu]}}
\newcommand{\Bdaphfreq}{\mathsf{B}^{[\Delta_{pq} t, \Delta_{pq}\nu]}}
\newcommand{\Dbdatime}{\mathsf{D}^{[\Delta_{pq} t]}}
\newcommand{\Dbdafreq}{\mathsf{D}^{[\Delta_{pq} \nu]}}
\newcommand{\Dbdaphtime}{\mathsf{D}^{[\Delta_{\alpha \beta} t]}}
\newcommand{\Dbdaphfreq}{\mathsf{D}^{[ \Delta_{\alpha \beta} \nu]}}
\newcommand{\Dbdlong}{\mathsf{D}^{[\Delta_{12} t, \Delta_{12} \nu]}}
\newcommand{\Dbdmedium}{\mathsf{D}^{[\Delta_{23} t, \Delta_{23} \nu]}}
\newcommand{\Dbdshort}{\mathsf{D}^{[\Delta_{34} t, \Delta_{34} \nu]}}
\newcommand{\Bdaphlong}{\mathsf{B}^{[\Delta_{12} t, \Delta_{12} \nu]}}
\newcommand{\Bdaphmedium}{\mathsf{B}^{[\Delta_{23} t, \Delta_{23} \nu]}}
\newcommand{\Bdaphshort}{\mathsf{B}^{[\Delta_{34} t, \Delta_{34} \nu]}}
\newcommand{\EDIT}[1]{\textcolor{black}{{#1}}}
\newcommand{\Sincc}{\textcolor{black}{\text{sinc}}}
\newcommand{\ATMNEW}[1]{\textcolor{black}{{#1}}}
\newcommand{\Bessel}{\textcolor{black}{\text{Bessel}}}
\newcommand{\LLA}{L1}
\newcommand{\LLB}{L2}
\newcommand{\LLC}{L3}
\newcommand{\EEE}{\bmath{\mathrm{D}}}
\newcommand{\GGG}{\bmath{\mathrm{G}}}
\section{Introduction and motivations}
\label{introduction}
A variety of new radio telescopes, precursors (e.g. ASKAP~\citep{johnston2008science}, MeerKAT~\citep{jonas2009meerkat}) and Pathfinders  (e.g. LOFAR~\citep{van2013lofar}, NenuFAR~\citep{zarka2015nenufar}) for the Square Kilometre Array~\citep{dewdney2009square} (SKA) \Refcom{are under} development or
used \Refcom{to image} wide field of view \Refnew{(FoV, i.e. the fractional portion of the primary beam at the full width at half maximum (FWHM))} sky surveys at high sensitivity, wide bandwidth and high spectral and temporal resolution. These radio telescopes produce an extremely
large volume of data, such that data storage and analysis are becoming more challenging for scientific research and engineering requirements e.g., to
transmit the data from the \Refcom{receivers} to the \Refcom{correlator} or in data reduction such as calibration and imaging. 
A typical example is the LOFAR telescope. Its $uv$-data (visibilities), assuming 24 core stations (excluding the remote and international stations) using 244 sub-bands with 64 channels  per sub-band, 4 hours observation time with a 1 s temporal resolution is predicted to 
be $\sim$8376 GB using the dual  high band antenna  (see LOFAR calculator\footnote{{\tt lofar.astron.nl/service/pages/storageCalculator\\/calculate.jsp}}).  However, observations with all the LOFAR national and international stations are capable of producing data volumes of the order of petabytes~\citep{sabater2017calibration}.
Survey capabilities with the future SKA  (unprecedented sensitivity, resolution and bandwidth) are expected to generate data by many orders of magnitude higher than any existing radio interferometer. 
This data volume will be \Refcom{even  larger} for  any SKA survey science that will integrate multiple beams and/or multiple phase tracking e.g., African Very Long
Baseline Interferometry (VLBI)  Network~\citep{gaylard2014african}, European VLBI Network (EVN)~\citep{keipema2015sfxc},  etc.
New techniques for data compression and storage systems must be 
developed for the transition from the current radio interferometers to the SKA.
Data compression is an advantageous solution for increasing the speed of the data transmission and to decrease the computational requirements for post-processing. Data compression also offers an alternative possibility for wide FoV observations because it offers significant reduction of \Refcom{the data volume} while preserving useful information to improve discovery and analysis accuracy.

\begin{figure*}
\centering
\includegraphics[width=.4\textwidth]{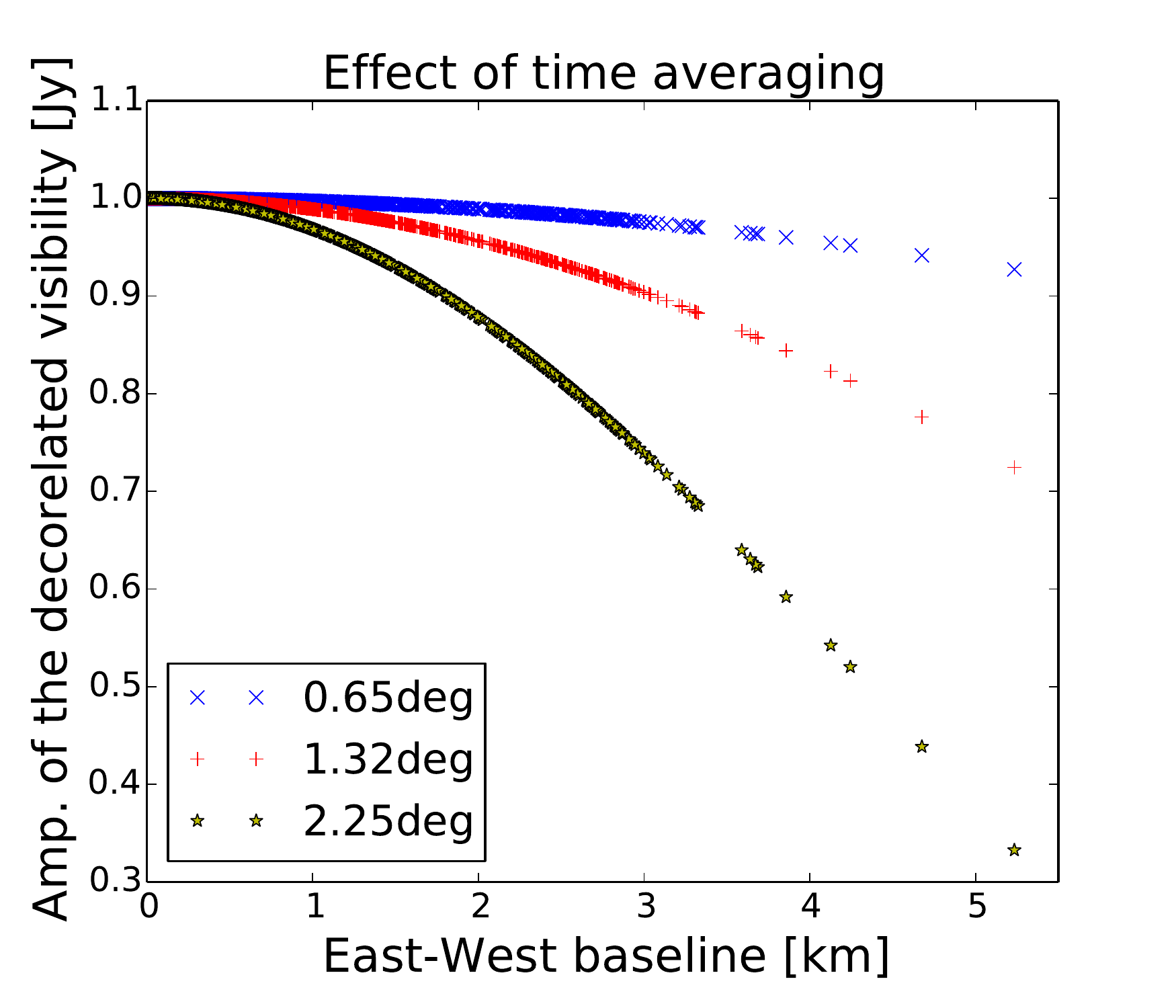}%
\includegraphics[width=.4\textwidth]{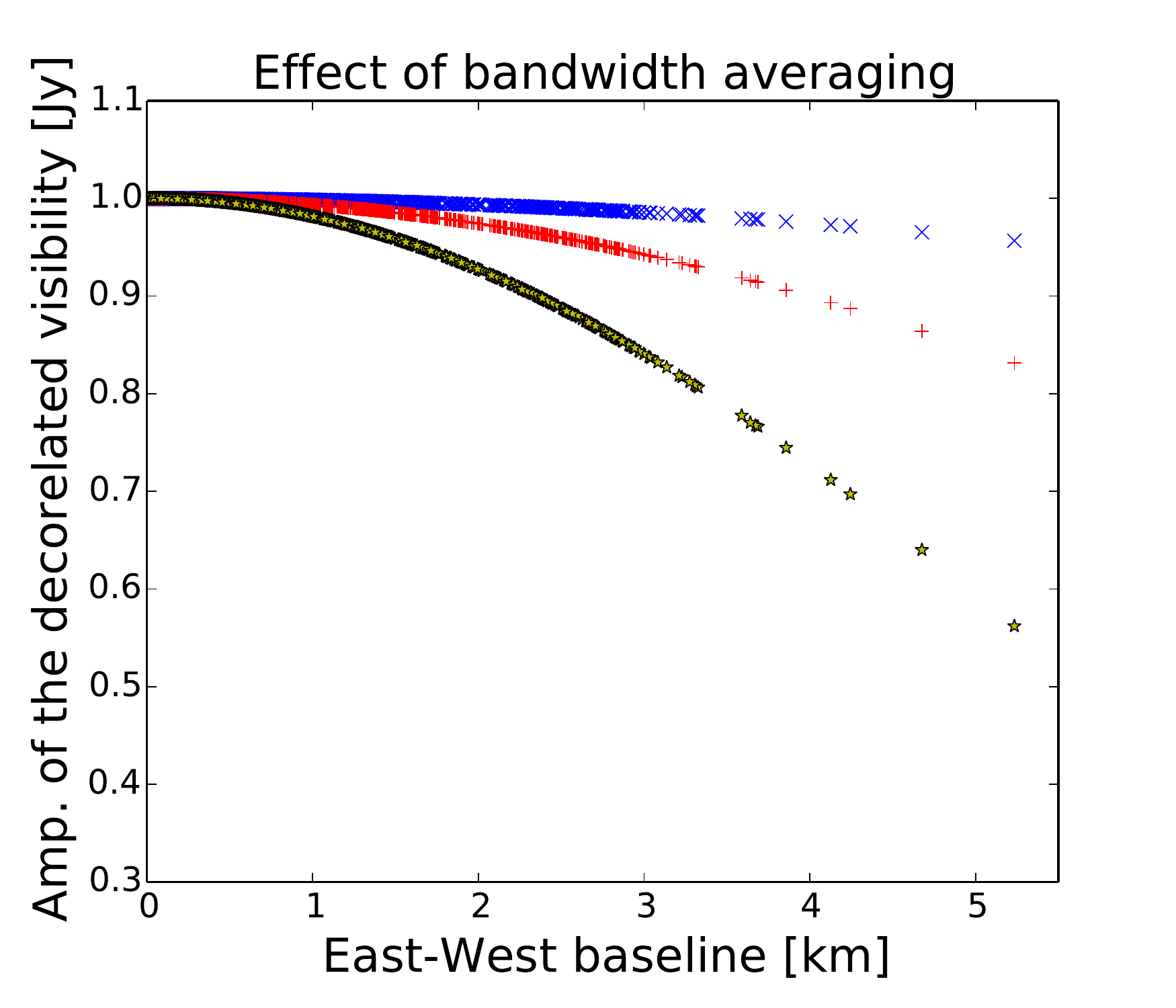}%
\caption{\ATMR{
 Amplitude loss: the apparent intensity of a 1 Jy source at 0.65 deg, 1.32 deg and 2.25 deg as seen by MeerKAT at 1.4 GHz 
as a function of East-West baseline components; \Refcom{(Left): data is \Refnew{simple} averaged across  15 s in time and frequency resolution is fixed to 84 kHz; 
(Right) data is \Refnew{simple} averaged across 0.84 MHz in frequency and time resolution is fixed to 1 s}. 
 }}\label{fig:srcat30arcmin_avg}
\end{figure*}

Traditionally,  radio interferometric correlators compress the visibility data by simply averaging the data, which may be averaged further in post-correlation to speed up processing. 
\Refcom{However, the challenge} in compressing the visibilities by simple averaging is that these visibilities decorrelate and the decorrelation is time-frequency dependent and baseline-dependent. The visibility from a baseline $pq$ \Refcom{(with vector  $\bmath{u}_{pq}=(u,v,w)$)} of a point source 
with brightness $S$ and coordinates $\bmath{l}=(l,m,n-1)$ is given by:
\begin{equation}
V_{pq} = S \exp \big \{-\ii \phi \big \},  ~ \phi(\bmath{u}_{pq}) = 2 \pi \bmath{u}_{pq}\cdot\bmath{l}.\label{eq:phase}
\end{equation}
\ATM{For sources with an increasing separation from the phase centre, the phase $\phi$ is increasingly large for a given baseline, and at some distance phase-wrapping within the averaging time-frequency will cause a strong decorrelation of the signal.}
%
\Refcom{Figure~\ref{fig:srcat30arcmin_avg}  is a simulated observation with MeerKAT at 1.4 GHz  showing the amplitude decorrelation  for a 1 Jy point source located at $0.65$ deg, $1.32$ deg  and $2.25$ deg  away from the phase tracking centre as a function of East-West baseline length.}
At this frequency, a MeerKAT survey must be able to image sources up to an angular distance of 0.65 deg (edge of the FoV at the FWHM of the primary beam (PB)) from the phase tracking centre with little to no smearing effects.
 But modern calibration and imaging techniques such as  MeqTrees~\citep{noordam2010meqtrees} or DDFacet~\citep{ctasseDDFacet} are able to correct for PB effects 
 \Refcom{far exceed the second sidelobe of the PB~\citep{mitra2015incorporation}}. 
 An accurate PB model is necessary for calibrating out the effects of the PB, and for improving image fidelity. A good PB model can significantly reduce artefacts in the image and improve its dynamic range, and an appropriate direction-dependent calibration procedure can further reduce artefacts and increase the \Refcom{dynamic range~\citep{mitra2015incorporation}.}
 \Refcom{Throughout  this paper, we use the term Field-of-Interest (FoI) to differentiate from the FoV when the region of interest to be imaged exceed \lq\lq the fractional portion of the PB at the FWHM’’.}
 The first  and the second null of the PB of MeerKAT at 1.4 GHz  fall at $\sim$1.32  and  $\sim$2.25 deg respectively.
\Refcom{In Figure~\ref{fig:srcat30arcmin_avg} the pre-averaged data is simulated using 1 s and 84 kHz for time and frequency resolutions respectively. To evaluate the time smearing the data is \Refnew{simple} averaged across 15 s
and the frequency resolution remains fixed to 84 kHz. Similarly, for the bandwidth smearing the time resolution is maintains to 1 s and the 
data is \Refnew{simple} averaged across 0.84 MHz in frequency. Results show that decorrelation/smearing is severe on  
longer East-West baselines than  shorter East-West baselines and that smearing is  a function of source position in the sky.}

\Refcom{Simple averaging could be used in a way to increase the signal-to-noise ratio (SNR)  within the FoI by suppressing the sidelobes from sources out of the FoI, but the drawback is that sources at the edges of the FoI will be smeared~\citep{lonsdale2004efficient,atemkeng2016}.
However, increasing the SNR based on averaging is feasible only if both the FoI and its edges are preserved from smearing, and sources out of the  FoI are suppressed. The later is resumed mathematically as follows:}
 
\begin{equation}
\mathrm{SNR} \approx \frac{S_\mathrm{smear}}{C_\mathrm{noise}+T_\mathrm{noise}}, \label{eq:snr}
\end{equation}
where $S_\mathrm{smear}$ is the signal of a source in the FoI \Refcom{(including the edges)} that \ATMR{must be preserved from smearing}, $C_\mathrm{noise}$ the signal from sources outside the FoI (i.e. confusion noise) that must be
subtracted from the FoI \ATMR{or must strongly decorrelate} and $T_\mathrm{noise}$  the thermal noise which is usually Gaussian and  intrinsic to the visibility measurement process. Ideally, one wants an increase in $S_\mathrm{smear}$  and a decrease in $C_\mathrm{noise}$ within the FoI, so that the overall SNR increased even if there is an increase  in $T_\mathrm{noise}$ in the case of weighted averaging.

\begin{figure*}
\centering
\includegraphics[width=.4\textwidth]{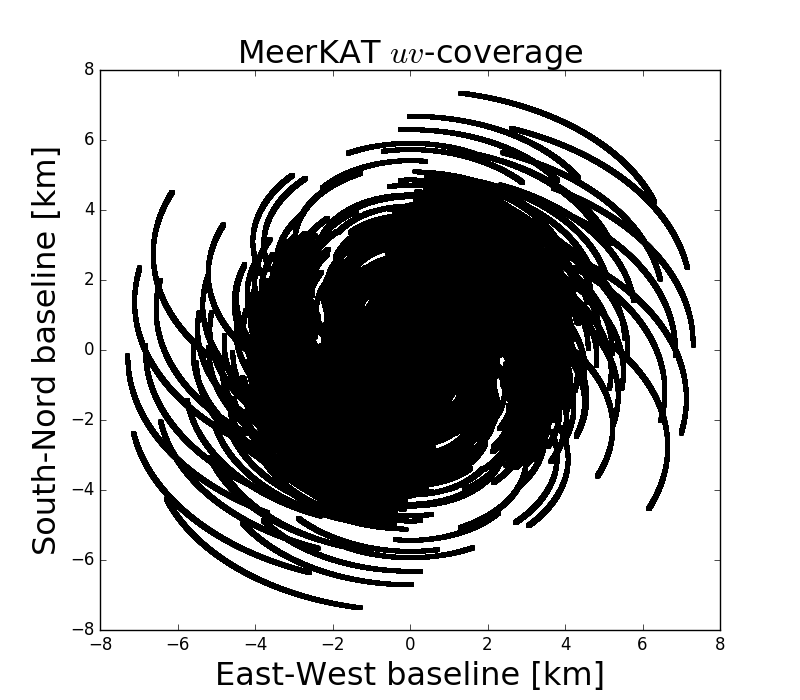}%
\includegraphics[width=.4\textwidth]{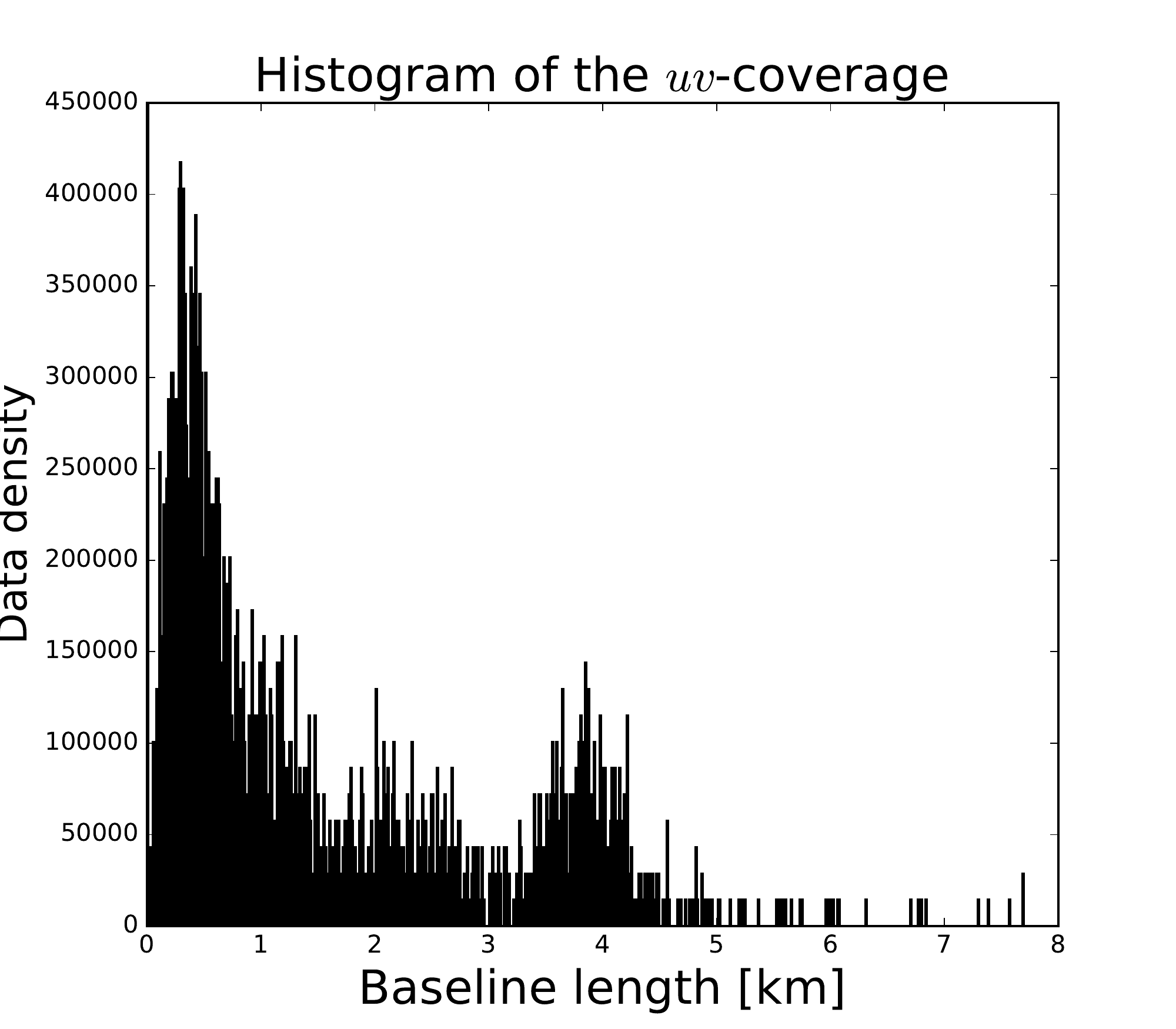}%
\caption{MeerKAT $uv$-coverage at 1.4 GHz and histogram depicting the data density as a function baselines length, \Refcom{4 hr} observation and 8 MHz bandwidth  showing clearly that the data \EDIT{are}  
condensed at the centre. \Refcom{Most of the data at the centre come from the short baselines.}}\label{fig:meerkat}
\end{figure*}

\Refcom{If the $uv$-coverage  of an interferometer is condensed at the centre then must of the data comes from the shorter baselines.  An example of this type of centrally condensed $uv$-coverage along with the $uv$-coverage histogram is illustrated in
Figure~\ref{fig:meerkat}.} The histogram shows the $uv$-coverage data density as a function of  effective baseline length.
 \EDIT{I}f more samples \EDIT{should be} \Refcom{averaged at the centre} and fewer  at the outer, 
decorrelation \EDIT{can be}  avoided
on the longer baselines and data compression \EDIT{would be} carried out on the shorter baselines. This method\EDIT{,} 
often referred to as baseline-dependent averaging (BDA),
was first proposed by \citet{cotton1989special,cotton1999special} as an approach for \EDIT{dealing with} wide field imaging with little to no bandwidth and time averaging effects. 

The idea of BDA is thus not novel, and has also been subject of discussion in many radio interferometry conferences, particularly \ATMR{the ability to use BDA for the SKA data processors.} 
\Refnew{\citet{atemkeng2016} discussed a baseline-dependent  window functions (BDWFs) scheme that has the effect to shape the FoI.  Several other techniques to shape the FoI using window functions have been proposed~\citep{lonsdale2004efficient,parsons2009calibration,parsons2016optimized}.
BDWFs are weighted-moving averaging of the  irregular sampled visibilities in the $uv$-space. The mathematical derivations for the
BDWFs show that the dirty image is the apparent sky multiplied  by the inverse Fourier transform of each  of the BDWFs.}
\ATMX{\Refnew{This work  removes the restriction of irregular sampling in $uv$-space adopted in \citet{atemkeng2016} and considers regular sampling and averaging in the $uv$-space as a BDA formalism.
To shape the FoI, the BDA formalism is applied to BDWFs, i.e. applying  weighted-moving averaging to the regular sampled visibilities in the $uv$-space. Throughout this paper, we will be referring BDA applied to BDWFs as BDAWFs.}
 Since an unweighted average represents theoretically maximum sensitivity at the centre of the FoI, a weighted averaging  will result in a loss in nominal sensitivity.  However, to alleviate the decrease in sensitivity, 
 BDWFs  are further extended by \citet{atemkeng2016},  showing that the use of overlapping BDWFs has the benefit of suppressing the far FoI sources compared to simple averaging, and could even recover some of the lost sensitivity while decreasing the overall far-field confusion noise. Overlapping BDWFs are sets of polyphase finite impulse response filters with order depending on the overlapping bins in the $uv$-space. The overlapping bins  compensate for the missing bins windowed with the BDWF.
We refer the reader to \citet{atemkeng2016} for an intensive discussion on BDWFs and properties of overlapping BDWFs.  \Refnew{The
mathematical framework derived from the BDAWFs
formalism shows that the dirty image is the apparent sky
multiplied by the inverse Fourier transform of a single BDWF.}}

\section{Mathematical background}
\label{sect:simpleavg}
We use the radio interferometry measurement equation (RIME) formalism, which
provides a model of a generic interferometer.
For details on the RIME formalism
see~\citet{hamaker1996understanding,smirnov2011revisiting2,smirnov2011revisiting}. \Refcom{In a single mathematical equation, the RIME
describes all the direction-dependent and direction-independent effects that may occur when an interferometric measurement is in process.}  
\Refcom{The 2-D Fourier transform full sky RIME\EDIT{,} following~\citet{smirnov2011revisiting2,smirnov2011revisiting}\EDIT{,} is given by:}
\begin{equation}
\VV_{pq} = \GGG_{pt\nu} \Big(\iint\limits_{lm}\EEE_{pt\nu} \II\EEE_{qt\nu}^\mathrm{H}\ee^{-\ii\phi}\Rd l \Rd m\Big)\GGG^\mathrm{H}_{qt\nu},\label{eq:rime:smirnov}
\end{equation}
where the superscript $(.)^\mathrm{H}$ denotes a Hermitian transpose operator. Here  a single visibility value  is denoted by $V_{pq}$ or in \Refnew{functional form} by $\VV_{pq}\equiv\VV(\bmath{u}_{pq})$ 
and the  sky distribution function by $\II\equiv\II(l,m)$.
The formalism groups  the product of \EDIT{direction-independent} Jones matrices corresponding to antenna $p$ into the matrix $\GGG_{pt\nu}$, and all its \EDIT{direction-dependent} effects 
into the matrix $\EEE_{pt\nu}$. 
We note that the PB pattern of each of the antenna that defines the directional sensitivity and the FoV of each of the antennas is part of the direction-dependent effects. 
The  term $\EEE_{pt\nu} \II\EEE_{qt\nu}^\mathrm{H}$ is the apparent
sky seen by baseline $pq$, and varies in time and frequency. \Refcom{For simplicity, throughout this work we assume
that both the sky and the direction-dependent gain  are invariant; therefore each of the baselines will  see the same apparent sky throughout the measurement process.}

Rotation of the  Earth causes the baseline phase to vary in time, and for multi-frequency observations 
the phase is constantly changing with time and frequency. 
\Refcom{In practical situations an  interferometer can only 
measure an average visibility over a fixed time-frequency lengths as  given by the \textit{sampling bin}:}
\begin{equation}
\Btf_{\RefEq{kr}} = \bigg [ t_k-\frac{\Delta t}{2},t_k+\frac{\Delta t}{2} \bigg ]
\times
\bigg [ \nu_r-\frac{\Delta\nu}{2},\nu_r+\frac{\Delta\nu}{2} \bigg ],  \label{eq:samplingbinnormal}
\end{equation}
where $\Delta t$ centered at $t_k$ and $\Delta \nu$ centred at $\nu_r$ are the sampling intervals in time and frequency respectively. \Refcom{The sampling bin  has two dimensions: the width and height  measured in time and frequency respectively.}
Let us denote $\VV_{pq}(\bmath{u}_{}(t,\nu))\equiv\VV(\bmath{u}_{pq}(t,\nu))$ as the ideal visibility distribution. After averaging in the correlator, the measured visibility becomes:
\begin{equation}
\VVM_{pq\RefEq{kr}} = \frac{1}{\Delta t \Delta \nu} 
\iint\limits_{\Btf_{\RefEq{kr}}}
\VV(\bmath{u}_{pq}(t,\nu))\Rd \nu \Rd t.
\label{eq2:conti}
\end{equation}
In the  time-frequency space the  bins are sampled equally on each baseline (assuming baseline-independent sampling), while in contrast in $uv$-space, they are not.
Ideally, all spatial frequencies up to the resolution of the longest baseline are
sampled in a 2-D continuous sky image. This requires Nyquist sampling of
the time-frequency space up to the highest spatial frequencies, corresponding to the longest baselines.
This is rarely possible because of the 
unsampled $uv$-space ``holes'' during an observation, the lower spatial frequency cut-off due to
physical element limitations and sampling bias in the low spatial frequency
region of the \Refnew{$uv$-space} compared to higher spatial frequencies due to
baseline distribution. For a fixed time-frequency length, a long baseline will cover a
longer track in $uv$-space compared to a shorter baseline, which results in
the \Refcom{lower Fourier modes} being oversampled compared to \Refcom{higher Fourier modes}. On
shorter baselines, the sampling bin  \Refcom{width and height are smaller}   compared to longer baselines; assuming baseline-dependent mapping. However, \Refnew{this work considers} two major sub-domains. (1) The correlator domain or the $t\nu$-space where the baselines are sampled equally onto a rectangular grid.  (2) The visibility domain or $uv$-space where the baselines are sampled differently and the overall data are mapped onto \Refcom{elliptical arcs/ribbons}.

Let us denote by $\Buv_{pq\RefEq{kr}}$ the matched $uv$-space sampling bin, which is baseline-dependent. The relation in Eq.~(\ref{eq2:conti}) can be rewritten  as:
\begin{equation}
\VVM_{pq\RefEq{kr}} = [ \VV_{pq} \circ \Ptf ](t_k,\nu_r) \text{, in } t\nu\text{-space or}\label{eq:avscontnuconvtv}
\end{equation}
\begin{equation}
\VVM_{pq\RefEq{kr}} = [ \VV_{pq} \circ \Puv_{pq\RefEq{kr}} ](\bmath{u}_{pq}(t_k,\nu_r)) \text{, in } uv\text{-space.}
\label{eq:avscontnuconvtvuv}
\end{equation}
\Refcom{Here $\circ$ stands for the convolution operator, and} $\Ptf$, $\Puv_{pq\RefEq{kr}}$ are normalised boxcar window functions 
defined in $t\nu$-space and $uv$-space respectively. The detailed derivations for these equations are developed in ~\citet{atemkeng2016}.
Eq.~(\ref{eq:avscontnuconvtv}) and~(\ref{eq:avscontnuconvtvuv}) are of importance because they clearly show  that 
visibility averaging  is equivalent to convolution at the centre of  the sampling bin of  the true visibilities and the boxcar window function. 
We emphasise that  the discussion above provides an alternative way to look at decorrelation/smearing. With averaging in effect, a useful mathematical model
may be of the following form:
\begin{equation}
\VVMF_{pq\RefEq{kr}} = \delta_{pq\RefEq{kr}} ( \VV\circ\Puv_{pq\RefEq{kr}} ),\label{eq:avscontnuconvtvuvdelta}
\end{equation} 
where $\delta_{pq\RefEq{kr}}$ \Refcom{denotes the Dirac delta functions i.e. a single nail sampling function.}
\subsection{Imaging}
To derive the effect of averaging on the image, we can reformulate  Eq.~(\ref{eq:avscontnuconvtvuvdelta})  as:
\begin{equation}
\VVMF_{pq\RefEq{kr}} = \FF\big\{\PP_{pq\RefEq{kr}}\big\} \bigg( \FF\big\{\II\big\} \circ \Puv_{pq\RefEq{kr}} \bigg),\label{eq:avscontnuconvtvuvdeltaim}
\end{equation} 
where the apparent sky $\II$ is the inverse Fourier transform of the ideal visibility measurement $\II=\FF^{-1}\big\{\VV\big\}$ and the point spread function $\PP_{pq\RefEq{kr}}$ is the inverse Fourier transform of the sampling function 
for the baseline $pq$ at  the discrete time-frequency bin $kr$, i.e. $\PP_{pq\RefEq{kr}}=\FF^{-1}\big\{\delta_{pq\RefEq{kr}}\big\}$.  Here $\FF^{}$ and $\FF^{-1}$ represent the Fourier transform and its inverse respectively.
Inverting the Fourier transform of the sum over all baselines of
Eq.~(\ref{eq:avscontnuconvtvuvdeltaim}) and sampling at each $\RefEq{kr}$ results in an  estimate 
of the sky image i.e. the ``dirty image'':
\begin{equation}
\label{eq:imaging}
\IID = \FF^{-1}\bigg\{ \sum_{pq\RefEq{kr}}W_{pq\RefEq{kr}} \VVMF_{pq\RefEq{kr}}\bigg\} ,
\end{equation}
where $W_{pq\RefEq{kr}}$ is the weight at the sampled point $pq\RefEq{kr}$; in all the extent of the $uv$-space
  $\WW=\sum_{pq\RefEq{kr}}W_{pq\RefEq{kr}}\delta_{pq\RefEq{kr}}$ in functional form, i.e. the weighted-sampling function.
Substituting Eq.~(\ref{eq:avscontnuconvtvuvdeltaim}) into  Eq.~(\ref{eq:imaging}) and applying the convolution theorem,
we now have:
\begin{equation}
\IID =  \sum_{pq\RefEq{kr}} W_{pq\RefEq{kr}} \PP_{pq\RefEq{kr}} \circ (\II\cdot\TT_{pq\RefEq{kr}}),\label{eq:dirtybdboxcar}
\end{equation}
with the apparent sky $\II$ now tapered by the baseline-dependent \emph{window response function} $\TT_{pq\RefEq{kr}}$, the latter being the inverse Fourier transform of the baseline-dependent boxcar window:
\begin{equation}
\TT_{pq\RefEq{kr}} = \FF^{-1}\Big\{ \Puv_{pq\RefEq{kr}} \Big\}.
\end{equation}
Interestingly, Eq.~(\ref{eq:dirtybdboxcar}) explicitly enforces conditions on the dirty image which has the dependence on all the individual image-plane response (IPR) tapers, $\TT_{pq\RefEq{kr}}$.
It should be noted that these IPR tapers are not completely arbitrary; in the sense that they depend on each  baseline length and orientation. 
\Refcom{Longer baselines have narrower IPR and are thus prone more to smearing than shorter baselines.}

In synthesis imaging, we assume that the sky is a constant signal (transient events are ignored), but a time
variable signal is measured because the projected baseline change in orientation and length as the Earth rotates. Also, the frequency coverage and array layout are used to fill in the synthesised aperture, making the signal depending on frequency and array layout. 
The boxcar window functions
 are linear but depend on baseline length, which varies with time and frequency: this is why in the entire $uv$-space, simple averaging is \textbf{not} a \textit{true-convolution} \Refcom{as demonstrated  in \citet{atemkeng2016}}. We   refer this as a ``\textit{pseudo-convolution}''. However, if one considers only a single East-West baseline, then simple averaging becomes a \textit{true-convolution} because the lengths of the boxcar window do not change along the $uv$-track. Simple averaging still remains a \textit{pseudo-convolution} for a baseline \Refcom{with a non-zero South-North component}.
When considering the entire $uv$-space then it is not sufficient to simply analyse the boxcar \Refnew{window functions}  IPRs. 
As opposed to true-convolution the pseudo-convolution is a linear time-frequency variant system, which leads to complexity in the analysis of the signal conditioning. 
\textcolor{black}{In practical situations, all the boxcar window functions are window-function-unweighted,
moving averages of the measured visibilities, rather than the ideal visibilities. Consider that $\Vs_{pqij}$ is the measured visibility sample at $pqij$ with high temporal and spectral resolution.  In this sense, we assume that $\Vs_{pqij}\equiv V_{pqij}$ if the noise term across all the visibility samples is ignored.  Averaging becomes a discrete convolution:}
\begin{equation}
\VVM_{pq\RefEq{kr}}= \frac{\displaystyle\sum\limits_{{i,j}\in \Bij_{\RefEq{kr}}} \Vs_{pqij} \Puv_{pq\RefEq{kr}}(\bmath{u}_{pqij}-\bmath{u}_{pq\RefEq{kr}})}
{\displaystyle\sum\limits_{{i,j}\in \Bij_{\RefEq{kr}}} \Puv_{pq\RefEq{kr}}(\bmath{u}_{pqij}-\bmath{u}_{pq\RefEq{kr}})},\label{eq:avscontnuconvtvuvdelta11}
\end{equation}
where the set $\Bij_{\RefEq{kr}}$ corresponds to the bin indices of the \Refcom{sampling bin}, \Refcom{i.e. $\Bij_{\RefEq{kr}} =  \{ ij:~t_i\nu_j \in \Btf_{\RefEq{kr}}   \}$}.

\Refcom{This work investigates an alternative approach for visibilities sampling, which emphasises that in the entire $uv$-space all the baselines should be regularly sampled then window function should be applied to shape the FoI. If the window function is a boxcar window or a BDWF then the regular sampling will results to an invariant window length in $uv$-space, which is now a \textit{true-convolution} in the entire $uv$-space as opposed to the work discussed in \citet{atemkeng2016}. A true-convolution in the entire $uv$-space means that in the $t\nu$-space,  the time-frequency sampling intervals now varies across baselines: longer sampling intervals on short baselines and shorter on long baselines. Using this novel approach, the sampling bin defined in Eq.~(\ref{eq:samplingbinnormal}) becomes baselines-dependent: the width and height of the sampling bin vary as a function of East-West baselines length. Also, with the novel approach the BDWFs in the $t\nu$-space are sampled equally but are changing in lengths and resolution across baselines. Each of these properties are shown in Figure~\ref{fig:bda-boxcar-uvleng-directtion}.
Interest in such techniques comes from the fact that:}
 \begin{itemize}
  \item There are some longer baselines where the data should be averaged more than some shorter baselines. This can be seen in the histogram of Figure~\ref{fig:meerkat}, where data are condensed for baseline lengths between $\sim$3.5 km and $\sim$4.2 km than some shorter baselines. These longer baselines have smaller East-West components and are less prone to decorrelation/smearing, and so the data should be averaged more.
  \item \ATMR{The \Refcom{sampling bin} for a single baseline \Refcom{with a non-zero East-West and South-North components}  should vary along the baseline $uv$-track depending on the baseline direction. \Refnew{This variation of the sampling bin should be taken into account for  regular sampling in the $uv$-space.}}
  \item The IPR taper for all the baselines may result in the same degree of decorrelation/smearing if the \Refcom{visibilities are regular sampled in the $uv$-space.}
  \item One may adapt signal processing methods that assume a \textit{true-convolution} to find the optimal matched IPR.  Finding an optimal matched IPR is beyond the scope of this paper, and part of an ongoing study.
\end{itemize}
\section{\Refcom{Baseline-dependent sampling and averaging: BDA}}
\label{sec:bda}
\subsection{ Effect on the image}
\label{sec:bdaeffectImage}
\begin{figure*}
\includegraphics[width=.25\textwidth]{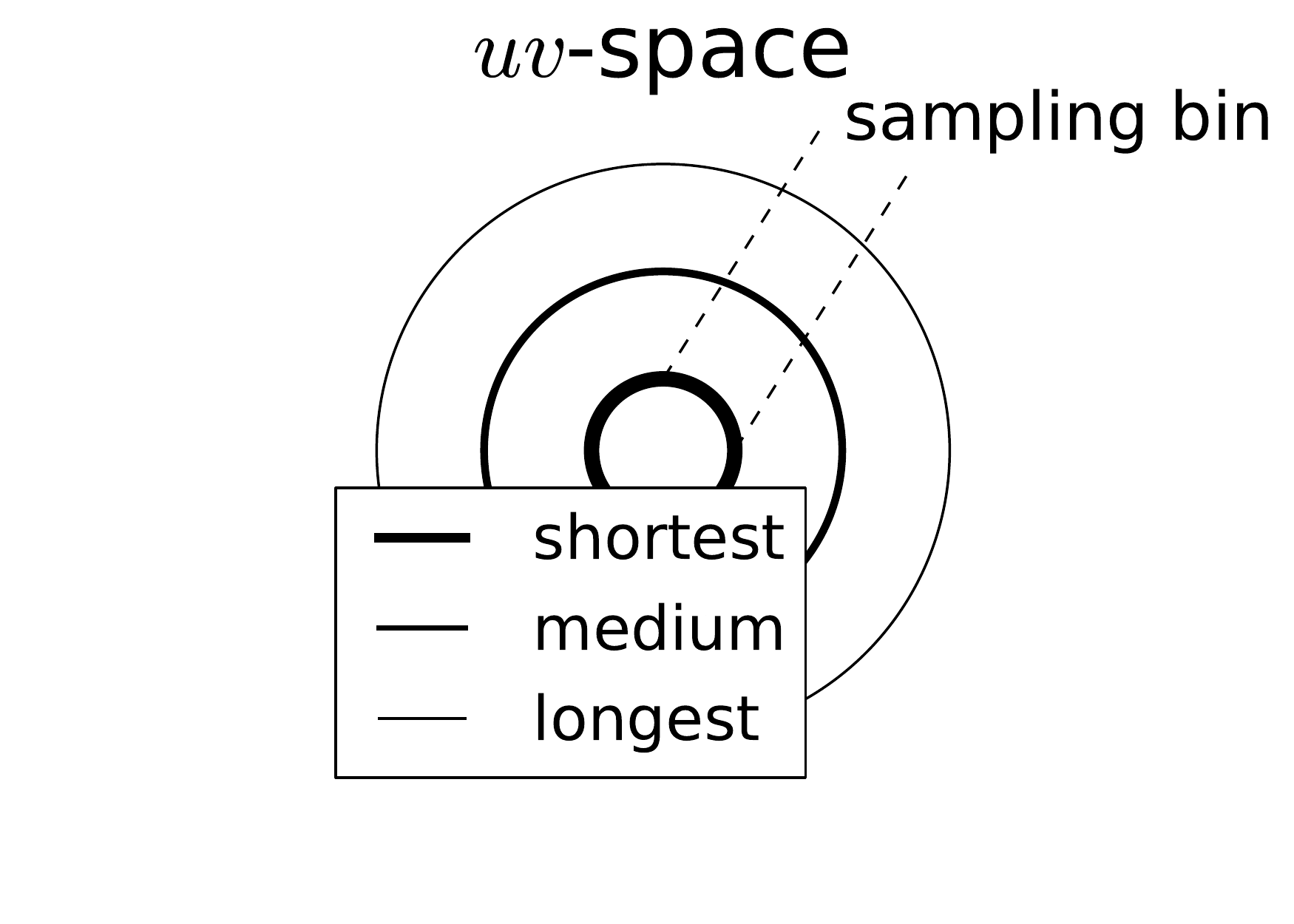}%
\includegraphics[width=.25\textwidth]{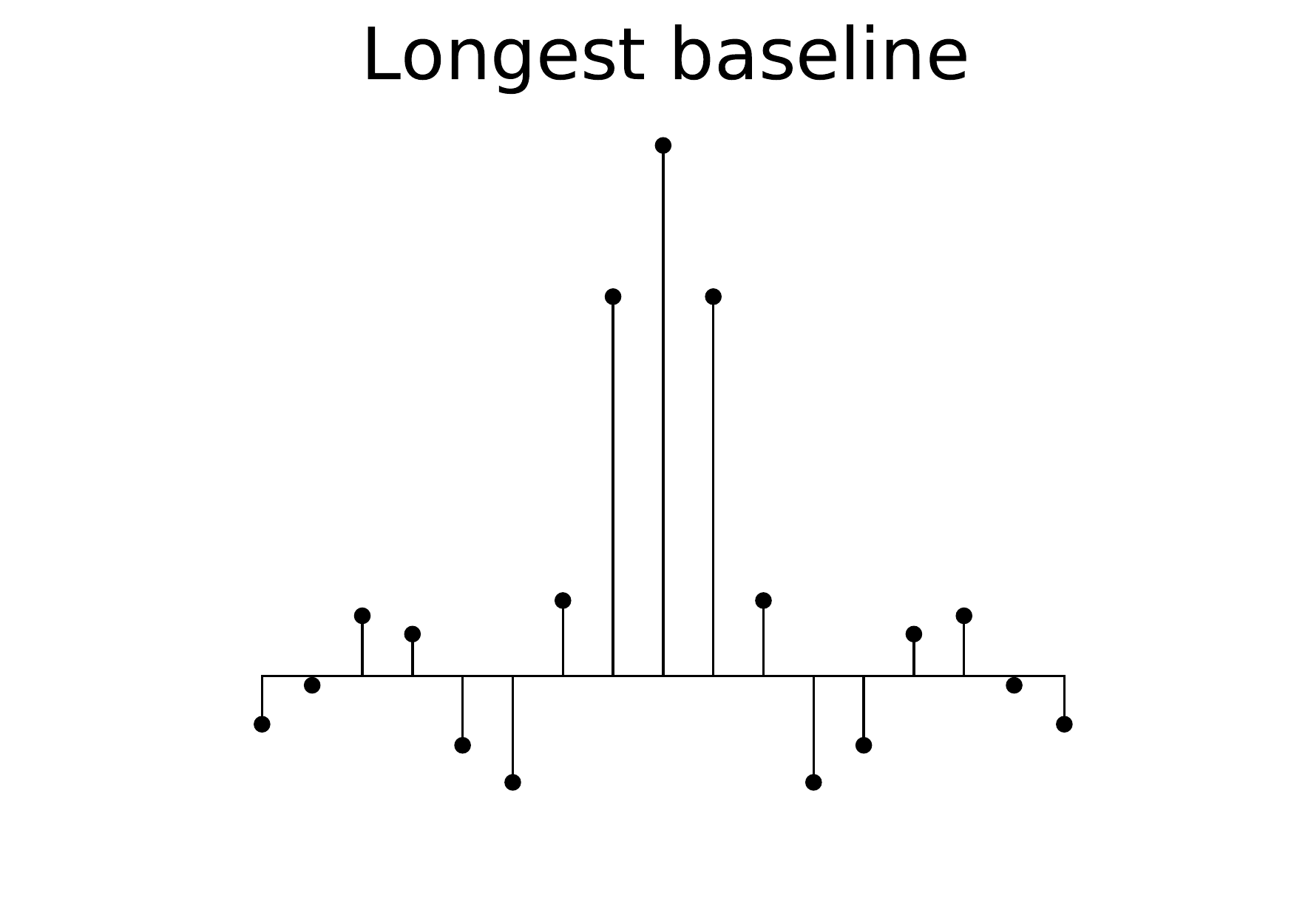}%
\includegraphics[width=.25\textwidth]{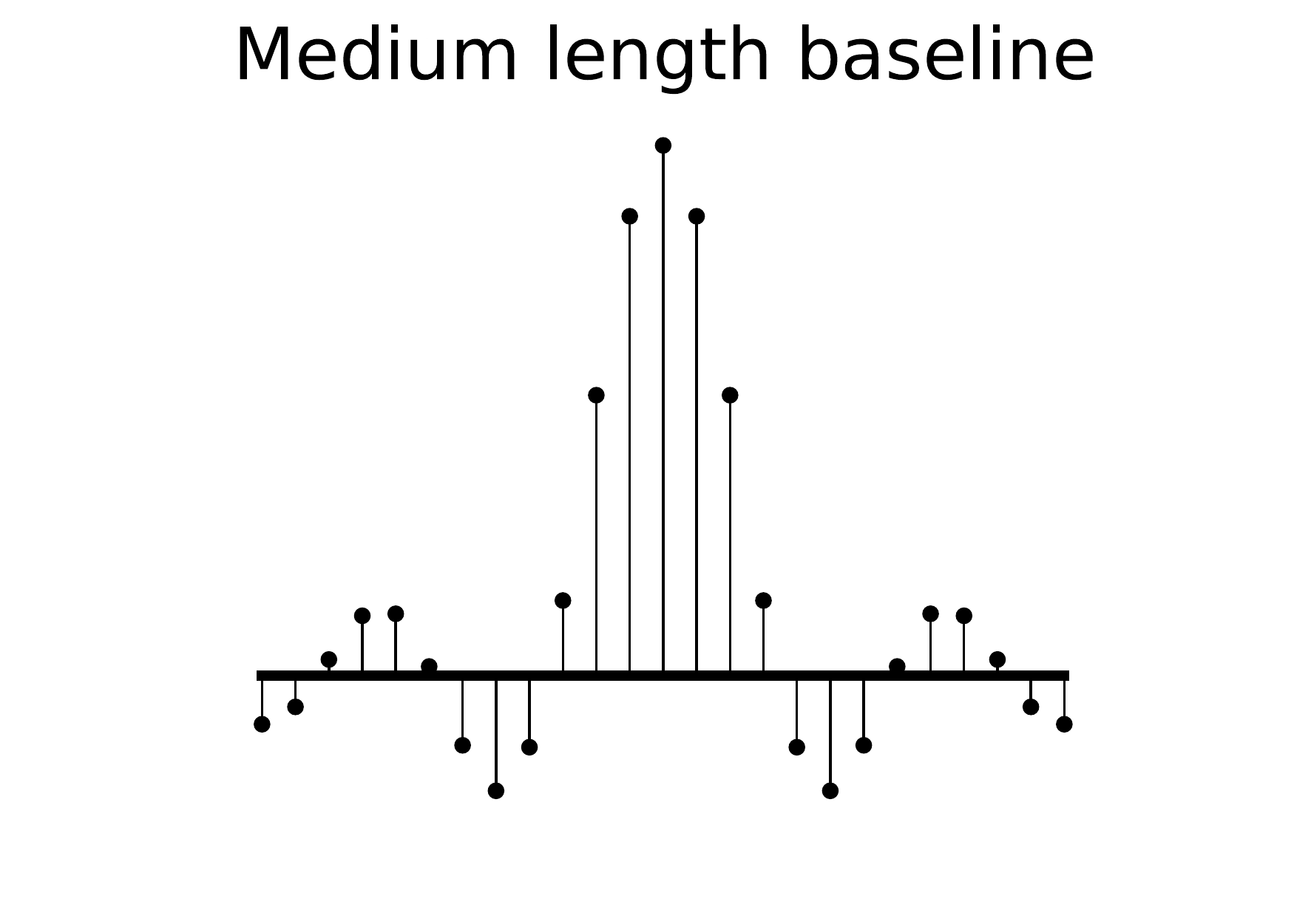}%
\includegraphics[width=.25\textwidth]{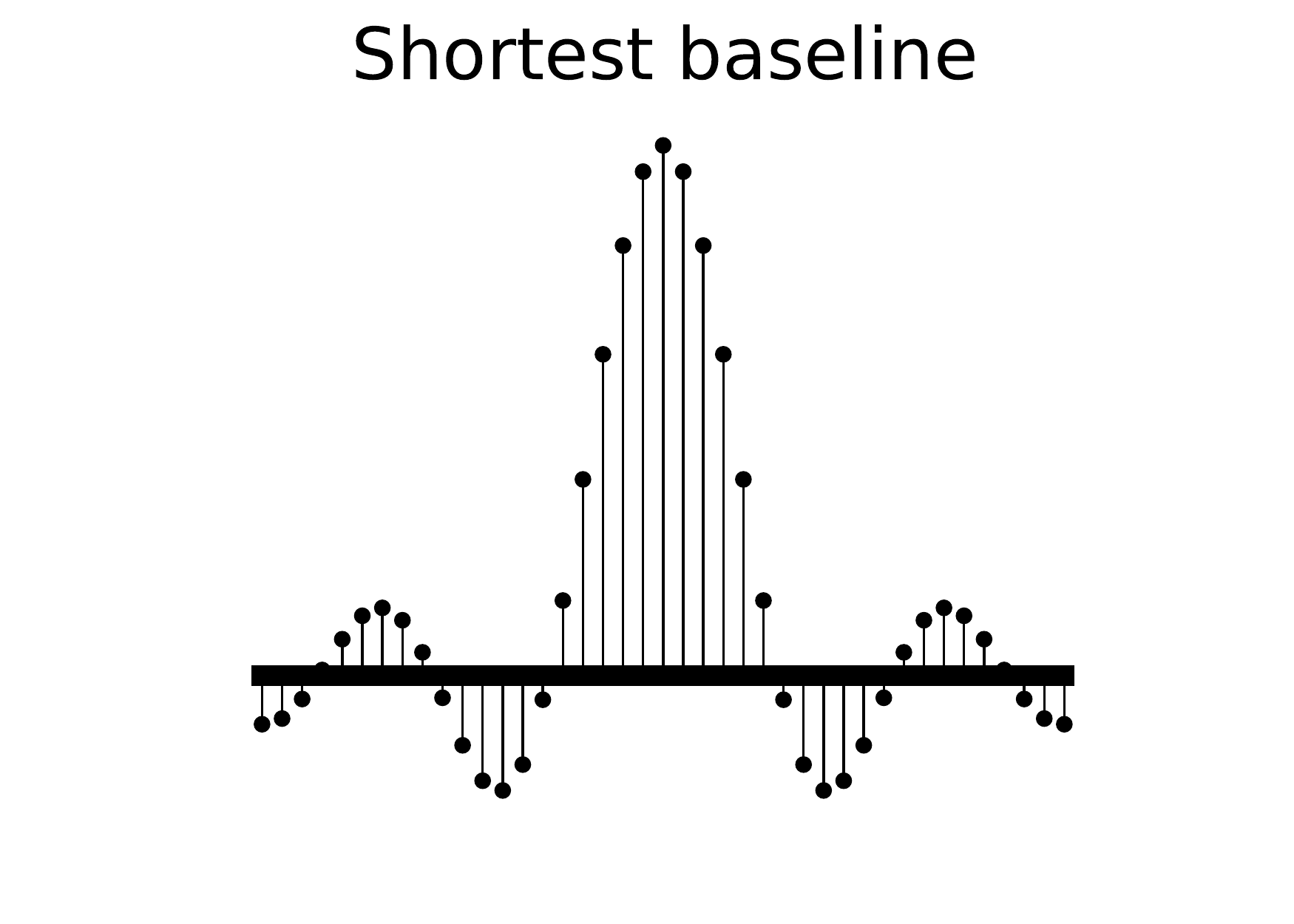}\\
\includegraphics[width=.25\textwidth]{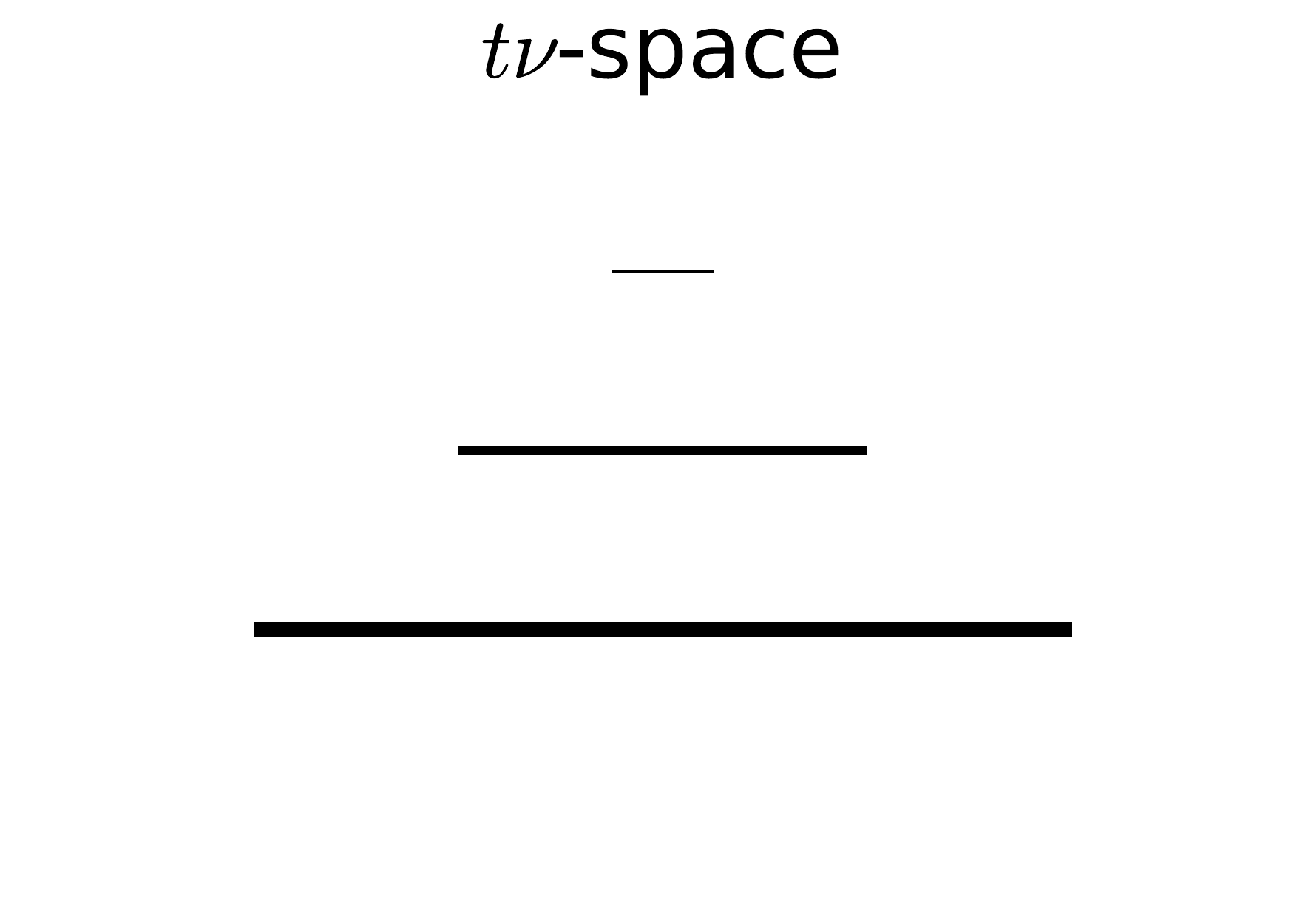}%
\includegraphics[width=.25\textwidth]{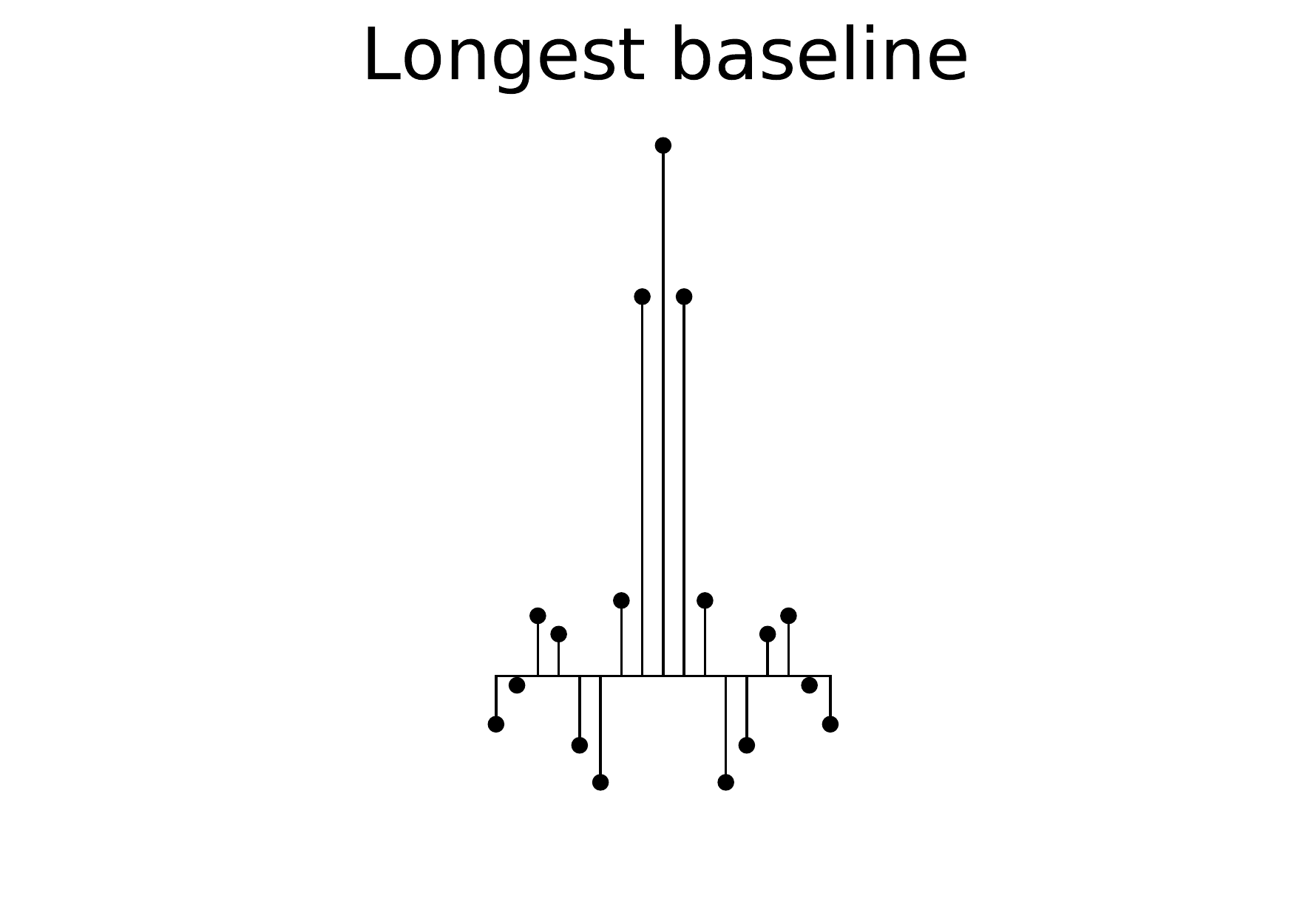}%
\includegraphics[width=.25\textwidth]{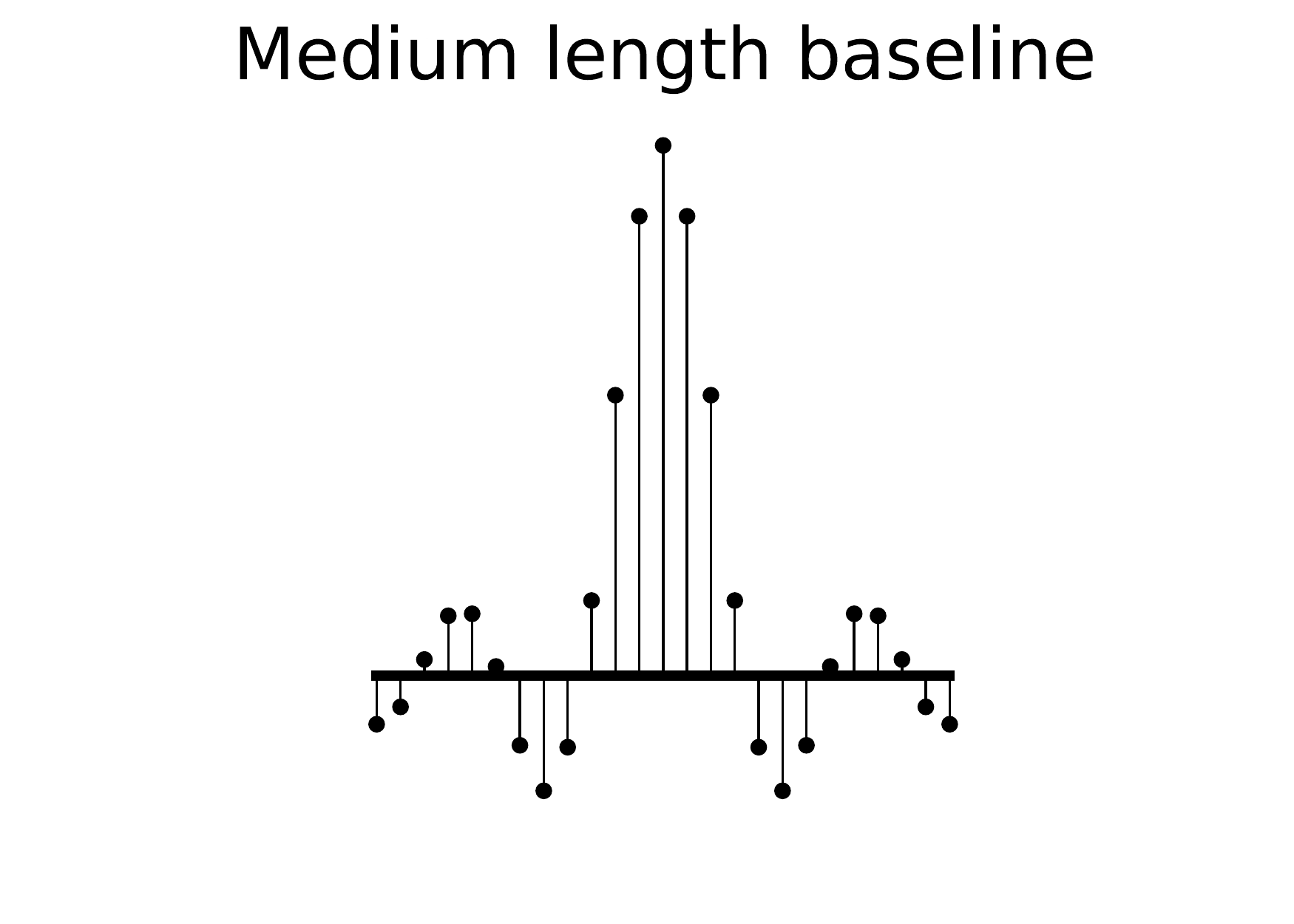}%
\includegraphics[width=.25\textwidth]{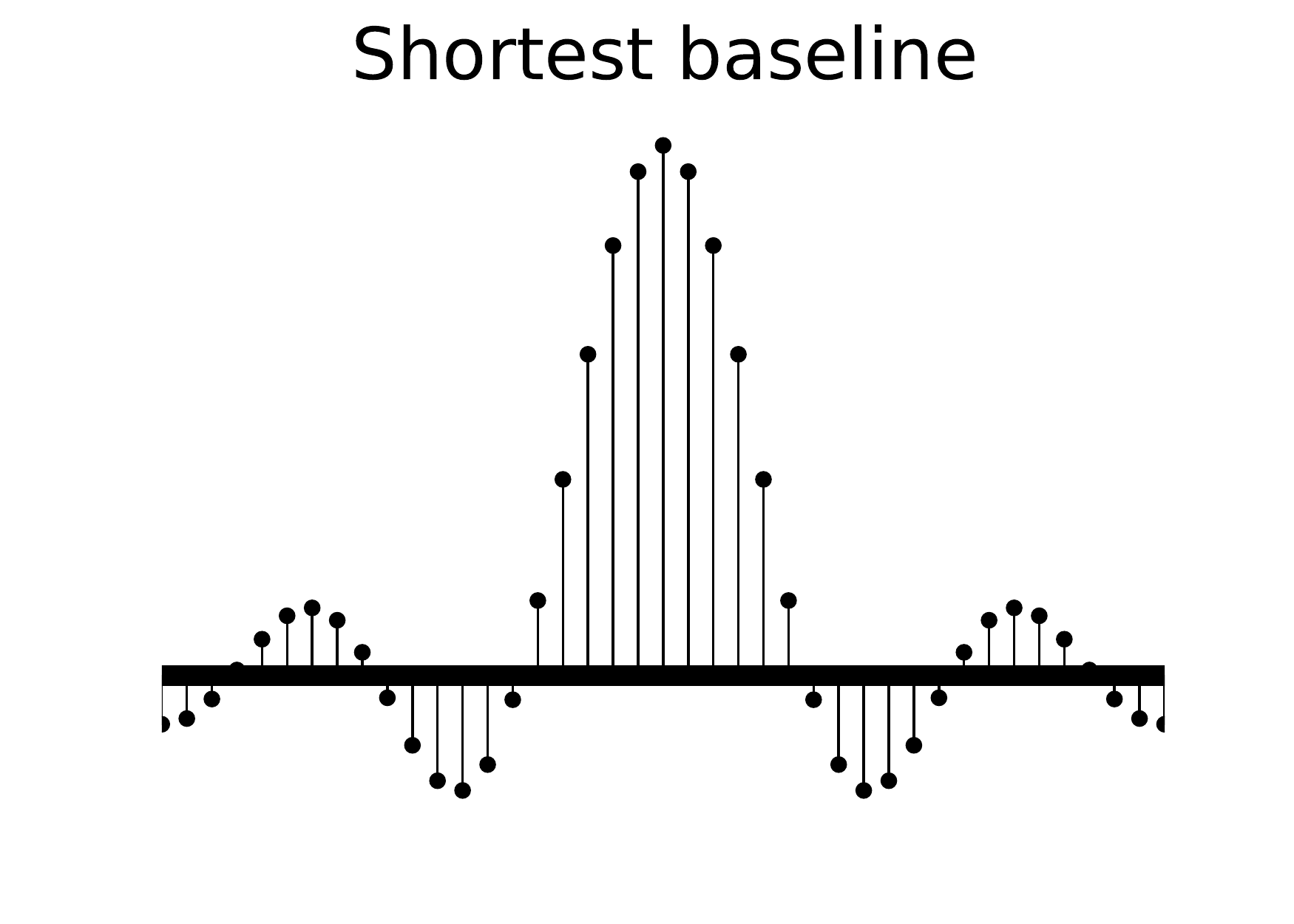}
\caption{
An East-West interferometer array: BDAWF defined in $\Refcom{uv}$-space (top) and in $t\nu$-space (bottom). In $uv$-space, the \Refcom{sampling bin}, \Refcom{the window resolution and  length}  remain constant across all the baselines, while the sampling rate varies with respect to the baseline length with shorter baselines oversampled and longer baselines downsampled. 
In $t\nu$-space, all the baselines are sampled equally but the \Refcom{sampling bin, window resolution and length  are now varying.}}
\label{fig:bda-boxcar-uvleng-directtion}
\end{figure*}
An interferometer measures the average visibility over a rectangular time-frequency bin given by $\Delta t$ and $\Delta \nu$\EDIT{:} this is the sampling bin
defined in Eq.~(\ref{eq:samplingbinnormal}). In $t\nu$-space, for a fixed length of \Refcom{time-frequency the corresponding sampling bin swept by different baselines in $uv$-space are not equal: shorter East-West baselines sweep smaller sampling bin and vice-versa.
Similarly, for a fixed sampling bin across all baselines in $uv$-space \Refnew{(baseline-independent sampling bin in $uv$-space)}, the corresponding time-frequency intervals in the $t\nu$-space vary with East-West baseline length: shorter time-frequency intervals  on long East-West baselines and longer time-frequency intervals on short East-West baselines.} \Refnew{Let us consider a baseline-independent sampling bin in $uv$-space} and let us denote the variant \Refnew{time and frequency} intervals by $\Delta_{\uu_{pq}} t$ and $\Delta_{\uu_{pq}} \nu$ \Refnew{ in $t\nu$-space respectively}.  
The \Refcom{sampling bin} becomes \Refnew{baseline-dependent in $t\nu$-space} (indicated here by the extra index $\uu_{pq}$, which is not found in Eq.~(\ref{eq:samplingbinnormal})):  
 \begin{alignat}{2}
\Bda_{\RefEq{kr}} =& \bigg [ t_k-\frac{\Delta_{\uu_{pq}} t}{2},t_k+\frac{\Delta_{\uu_{pq}} t}{2} \bigg ]\nonumber\\ 
&\times
\bigg [ \nu_r-\frac{\Delta_{\uu_{pq}}\nu}{2},\nu_r+\frac{\Delta_{\uu_{pq}}\nu}{2} \bigg ].
\end{alignat}
Figure~\ref{fig:bda-boxcar-uvleng-directtion} shows a typical  \Refnew{baseline-independent sampling bin in $uv$-space} (top-left) and baseline-dependent  \Refcom{sampling bin} \Refnew{in $t\nu$-space (bottom-left).}
\Refnew{If we denote in \Refnew{function form} by 
$\mathcal{D}$   the area of the baseline-independent sampling bin} in $uv$-space then we have:
\begin{eqnarray*}
 \mathcal{D}: \Bda &\rightarrow& \mathbf{\mathcal{R}}\\
                   t,\nu &\mapsto& d_{\uu_{pq\RefEq{kr}}},
\end{eqnarray*}
where $\mathbf{\mathcal{R}}$ is the set of real numbers. \Refcom{One can decomposed  $d_{\uu_{pq\RefEq{kr}}}$ as the product of the width $d_{\uu_{pqk}}$ and height $d_{\uu_{\RefEq{pqr}}}$ of the  sampling bin}:
\begin{alignat}{2}
 d_{\uu_{pq\RefEq{kr}}}&=d_{\uu_{pqk}}\times d_{\uu_{\RefEq{pqr}}}.
\end{alignat}
For 
$(t_i,\nu_j)\neq (t_k,\nu_r)$, \Refcom{  $d_{\uu_{pqk}}$ and  $d_{\uu_{\RefEq{pqr}}}$} are 
given by:
\newcommand{\D}{\mathcal{D}}
\newcommand{\Dim}{\mathcal{D}\mathrm{im}}
\begin{equation}
d_{\uu_{pqk}}=\sum_{t_i\nu_j}\norm{\bmath{u}_{pq}(t_i-t_k, \nu)},
\end{equation}
\begin{equation}
d_{\uu_{\RefEq{pqr}}}=\sum_{t_i\nu_j}\norm{\bmath{u}_{pq}(t, \nu_j-\nu_r)}\EDIT{,}
\end{equation}
\Refcom{where $t_i\nu_j\in\Bda_{\RefEq{kr}}$.} 
\Refcom{ If the visibilities are regular sampled along all the baselines in the $uv$-space then for all East-West baselines $\alpha \beta \neq pq$ with $\norm{\bmath{u}_{\alpha \beta}}\neq \norm{\bmath{u}_{pq}}$ 
the following constraints  must be satisfied:}
\begin{alignat}{2}
d_{\uu_{\alpha \beta k}}&=d_{\uu_{pq k}}~ \mathrm{and}~ d_{\uu_{ \RefEq{\alpha\beta r}}}&=d_{\uu_{\RefEq{pqr}}}.
\end{alignat}

Let us \EDIT{see} what Eq.~(\ref{eq:dirtybdboxcar}) becomes in the case \Refcom{of regular sampling along all the baselines in $uv$-space.}
The $uv$-space boxcar window\EDIT{,} $\Puv_{pq\RefEq{kr}}$ is now approximately equal in length across all East-West baselines, i.e. for all East-West baselines $\alpha \beta \neq pq$:
\begin{equation}
\Puv_{\alpha\beta \RefEq{kr}} \approx \Puv_{pq\RefEq{kr}}.\label{eq:bdabox}
\end{equation}
\Refcom{Does this meant that $\TT_{\alpha\beta \RefEq{kr}} \approx \TT_{pq\RefEq{kr}}$?} The latter will be always true in theory and not  in practice. \Refcom{Note that} while the length of the boxcar window is equal for all baselines in $uv$-space, the  boxcar window is sampled differently (the top panel of Figure~\ref{fig:bda-boxcar-uvleng-directtion} illustrates this in the case where the boxcar window is replaced with a sinc-like window).
The boxcar window  is \Refnew{downsampled} on the longer East-West baselines and \Refnew{oversampled} on the shorter East-West baselines, which then results to $\TT_{\alpha\beta \RefEq{kr}} \neq \TT_{pq\RefEq{kr}}$.
However, 
if the pre-averaged visibilities are sampled at  significantly  higher temporal and
spectral resolution (at the cost of computation) then one can 
assume that all these boxcar windows at different baselines are sampled equally.
Considering this assumption,  we can write:
\begin{equation}
\TT_{pq\RefEq{kr}}\approx \TT_{\alpha\beta \RefEq{kr}}\EDIT{.}\label{eq:bda2taper}
\end{equation}
Eq.~(\ref{eq:dirtybdboxcar}) becomes:
\begin{equation}
\IID  \approx \sum_{pq\RefEq{kr}} W_{pq\RefEq{kr}} \PP_{pq\RefEq{kr}} \circ \II\cdot\TT_{}, \label{eq:bda2}
\end{equation}
where $\TT_{}= \TT_{pq\RefEq{kr}}\approx \TT_{\alpha\beta \RefEq{kr}}$ is the smearing response, which is
now the effect of a single taper on the image. One can  summarise Eq.~\Refcom{(}\ref{eq:bda2}\Refcom{)} as:
\begin{equation}
\IID  \approx \IIA\cdot\TT_{}, \label{eq:bda2sumarrise}
\end{equation}
where $\IIA$ is the apparent image corrupted by all the effects that affect the signal from the source to the measurement and noise. 
\Refnew{The result in Eq.~(\ref{eq:bda2sumarrise}) is one of the  mathematical derivation achieved in this work, which shows that with BDA or BDAWFs in effect, the dirty image is the apparent sky multiplied by a single taper.}

\subsection{Implementation with current \Refcom{storage schemes}}
\label{BDA:impl}
In practice, most existing software implementations assume that the correlation matrix is a regular grid in time and frequency.
Averaging entries in this correlation matrix
over long times for short baselines and short times for long baselines results in an irregular grid.  
A better idea is to map this irregular grid onto a correlation matrix (i.e. regular grid) by either
flagging out the supplementary points, or duplicating the averaged values onto these supplementary points.

{\textit{Flagging:}}
Most of the radio interferometric data reduction software  has a flagging capability, \EDIT{through which} bad data can be flagged and ignored. For BDA, we exploit this capability to force interferometric data reduction software to ignore some entries of the regularly gridded plane (e.g. the correlation matrix).
In the flagging procedure, one has to make sure  that the \Refcom{sampling bin}  contains an 
odd number of data points in time as well as in frequency.
This condition must be verified on all baselines otherwise the average
baseline vector may not coincide with the mid-time and \Refnew{mid-}frequency vector and \EDIT{this} could lead to a phase shift. 
If this condition is satisfied, the average value is assigned to the midpoint of the \Refcom{sampling bin}. The other entries of the \Refcom{sampling} bin are flagged. 
This flag will cause missing samples to be ignored during post-processing. 

{\textit{Duplication:}}
This method consists of duplicating the average value at all entries of the \Refcom{sampling bin in $t\nu$-space}. While this process is easier to implement than the flagging method, it may not serve the purpose of data compression and/or quick computation for post-processing. It is easier to implement in the sense that one may not care or always verify that the
number of visibilit\EDIT{y} points in the \Refcom{sampling bin} is an odd number. Furthermore, the data size of the resulting
data set remains the same as the pre-averaged data set, \EDIT{since} all values are duplicated along the pre-averaged data set.
This method may be used in practice for cases where one does not want to estimate the averaged $uv$-coordinates from
the pre-averaged data set. 

{\textit{Semi-duplication and flagging:}}
This method consists of combining the flagging and the duplicate methods in order to benefit from their full advantages.
 \EDIT{In} so doing, we seek both data compression and quick computation\EDIT{,} while making implementation easier to handle.
The idea is to duplicate the averaged bin along two central entries of the \Refcom{sampling bin} if the total number of entries within this \Refcom{sampling bin} is even\EDIT{,} otherwise, the averaged bin is assigned only \EDIT{to} the central \Refnew{point} of the \Refcom{sampling}
bin. Any other entry is then flagged.

\subsection{Compression and computation}
The compression factor is defined as the ratio between the sizes of the pre-averaged  (high-res) data and the averaged  (low-res) data. In terms of the number of visibility samples, the high-res data size  
is:
\begin{alignat}{2}
N_\mathrm{vis}^\mathrm{hires}   &= N_{\mathrm{bl}}\times N_\mathrm{sub} \times N_\mathrm{pol} \times N_{t}^\mathrm{hires} \times N_{\nu}^\mathrm{hires}
,\label{eq:memorychap4hires}
\end{alignat}
where  $N_{\mathrm{bl}}$ is the number of baselines, $N_\mathrm{sub}$ the number of 
sub-bands, $N_\mathrm{pol}$ the number of polarisation, $N_{t}^\mathrm{hires}$ and $N_{\nu}^\mathrm{hires}$ the number
of timeslots and channels of the high-res data  respectively. 
\Refcom{For $n_{pq\RefEq{kr}}=n_{pqk}\times n_{\RefEq{pqr}}$ number of samples in the \Refcom{sampling} bin for 
a given baseline $pq$, with  $n_{pqk}$ and $n_{\RefEq{pqr}}$ the baseline number of time and frequency samples respectively.}
\Refcom{If one were to adopt a new storage scheme for BDA where there is no flagging or duplicated visibility samples, the  data size  \textcolor{black}{in terms of number of visibility samples} will be}:
\begin{alignat}{2}
 N_\mathrm{vis}^\mathrm{\Refcom{\scalebox{0.5}{BDA}}} &=\sum_{pq\RefEq{kr}} N_\mathrm{sub} \times N_\mathrm{pol} \times \frac{N_{t}^\mathrm{hires}\times N_{\nu}^\mathrm{hires}}{n_{pqk}\times n_{\RefEq{pqr}}}.\label{eq:memorychap4bd}
\end{alignat}
The compression factor after simplifications is then:
\begin{alignat}{2}
\mathrm{CF}_\mathrm{}  &=\frac{N_\mathrm{vis}^\mathrm{hires}}{N_\mathrm{vis}^\mathrm{\Refcom{\scalebox{0.5}{BDA}}}}
	  &=N_\mathrm{{bl}}\times \Bigg(\sum_{pq\RefEq{kr}} \frac{1}{n_{pqk}\times n_{\RefEq{pqr}}}\Bigg)^{-1}\label{compresionbdafactorx}.
\end{alignat} 
In the case of simple averaging  $n_{pqk}=n_{t}$, $n_{\RefEq{pqr}}=n_{\nu}$
with  $n_t$ and  $n_{\nu}$ the number of time  and frequency samples  averaged on each of the baselines. After simplifying Eq.~(\ref{compresionbdafactorx}) we have:
\begin{alignat}{2}
\mathrm{CF}_\mathrm{} &=n_t\times n_{\nu}.
\end{alignat} 
In the following sections, we refer to \EDIT{the} compression factor as \CF{$\mathrm{CF}_t$}{$\mathrm{CF}_{\nu}$}, where $\mathrm{CF}_t$ and $\mathrm{CF}_{\nu}$ are the compression factors in time 
and  frequency for the interferometer array respectively. The notations \CF{$\mathrm{CF}_t$}{$1$} and  \CF{$1$}{$\mathrm{CF}_{\nu}$} imply that the data \EDIT{are} 
compressed only in time by a factor of $\mathrm{CF}_{t}$ and only in frequency
 by a factor of $\mathrm{CF}_{\nu}$ respectively.
 \textcolor{black}{\Refcom{For BDA formalism,}  the shorter baselines are compressed by much more than $\mathrm{CF}$ and  the longer baselines
 by much less\EDIT{, while this corresponds}  to $\mathrm{CF}$ for the interferometer overall compression factor, which remains constant for all the baselines with simple averaging.}

\ATMR{The computational cost  $C^\mathrm{cost}$ during the compression  of the overall data for an individual interferometer  remains equivalent for  both BDA and simple averaging if their resulting compressed data are of the same size. The compression cost will scale as:}
\begin{alignat}{2}
 C^\mathrm{cost}&\sim \mathcal{O} (N_\mathrm{vis}^\mathrm{\Refcom{\scalebox{0.5}{BDA}}}\mathrm{CF})\\
	&\sim\mathcal{O} (N_\mathrm{bl}N_v \mathrm{CF})\\
	&\sim\mathcal{O} (N_\mathrm{bl}N_v n_tn_{\nu}),
\end{alignat}
 where $N_v$ is the number of visibilities and  $ \mathcal{O} (N_v n_t n_{\nu})$ the compression cost on each individual baseline after simple averaging respectively. But note that on each individual baseline the cost $C^{\mathrm{cost}}_{pq}$ then varies for BDA which scale as:
 \begin{alignat}{2}
  C_{pq}^\mathrm{cost}&\sim \mathcal{O} (N_{pqv} n_{pqk} n_{\RefEq{\RefEq{pqr}}}),
 \end{alignat}
 with $N_{pqv}$  the baseline-dependent number of resulting visibilities on $pq$ after BDA.
For shorter baselines $C_{pq}^\mathrm{cost}\ll \mathcal{O} (N_v n_t n_{\nu})$ while on the longer baselines $C_{pq}^\mathrm{cost}\gg \mathcal{O} (N_v n_t n_{\nu})$ but the overall computation cost leads to:
\begin{alignat}{2}
 C^\mathrm{cost}&\sim\mathcal{O}\Big(\sum_{pq\RefEq{kr}} N_{pqv} n_{pqk} n_{\RefEq{pqr}}\Big)\\
 &\sim \mathcal{O} \big(N_\mathrm{bl}N_v n_t n_{\nu}\big).
\end{alignat}
\subsection{Noise and noise penalty}
\label{sect:noisepenalty}
Let us look at what the estimates theoretical thermal noise induced by BDA become in each of the averaged  visibilities.
If for the high-res data, we assume that the noise term has constant r.m.s $\sigma_\mathrm{s}$ across all the baselines, then the noise induced  in each of the \Refcom{BDA} visibility is given by:
\begin{equation}
\sigma_{pq\RefEq{kr},\Refcom{\scalebox{0.5}{BDA}}}^2 = \frac{1}{n_{pq\RefEq{kr}}^2 } \sum_{i=1}^{n_{pq\RefEq{kr}}} \sigma_\mathrm{s}^2  = \frac{\sigma_\mathrm{s}^2}{n_{pq\RefEq{kr}}}\label{noise:avgbin}.
\end{equation}
Let us assume that the noise is uncorrelated across averaged visibilities.
The average of the squared error norm in each pixel of the dirty  image is then:
\begin{equation}
\sigma_{pix,\Refcom{\scalebox{0.5}{BDA}}}^2 = \frac{ (\sum_{pq\RefEq{kr}} W_{pq\RefEq{kr}}^2 \sigma_{pq\RefEq{kr},\Refcom{\scalebox{0.5}{BDA}}}^2) }{ (\sum_{pq\RefEq{kr}} W_{pq\RefEq{kr}})^2 },
\end{equation}
which for natural image weighting $W\equiv1$ simplifies to:
\begin{equation}
\sigma_{pix,\Refcom{\scalebox{0.5}{BDA}}}^2 = \Bigg(\frac{\mathrm{CF}\sigma_\mathrm{s}}{N_\mathrm{vis}^\mathrm{hires}}\Bigg)^2\sum_{pq\RefEq{kr}} \frac{1}{n_{pq\RefEq{kr}}}\label{noise:bda_pixel}.
\end{equation}
It is clear that the noise induced by BDA is completely different across baseline visibility samples because the number of averaged samples  are quite different; this is  expected from Eq.~(\ref{noise:avgbin}). In the case of simple averaging, Eq.~(\ref{noise:bda_pixel}) is reduced to:
\begin{alignat}{2}
\sigma_{pix,\Refcom{\scalebox{0.5}{AVG}}}^{2} &= \frac{\mathrm{CF}}{N_\mathrm{vis}^\mathrm{hires}n_tn_{\nu}}\sigma_{\mathrm{s}}^2\\
	      &=\frac{1}{N_\mathrm{vis}^\mathrm{hires}}\sigma_{\mathrm{s}}^2 
	      =\frac{1}{N_\mathrm{vis}^\mathrm{\Refcom{\scalebox{0.5}{AVG}}}n_tn_{\nu}}\sigma_{\mathrm{s}}^2,\label{noise:bda_pixelx1}
\end{alignat}
where $N_\mathrm{vis}^\mathrm{\Refcom{\scalebox{0.5}{AVG}}}$ is the number of visibilities in the simple averaged data, \Refnew{the index AVG stands for simple averaging}. Refer to Appendix~A for a detailed proof of Eq.~(\ref{noise:bda_pixel}) and~(\ref{noise:bda_pixelx1}). The derivation in Eq.~(\ref{noise:bda_pixelx1}) matches the result of the mathematical expectation of the squared error
norm in each pixel of the dirty image in the case of simple averaged as shown in \citet{atemkeng2016}.
\ATMR{It is  clearly shown in Eq.~(\ref{noise:bda_pixelx1}) that  $\sigma_{pix,\Refcom{\scalebox{0.5}{BDA}}}=\sigma_{pix,\Refcom{\scalebox{0.5}{AVG}}}$. 
 Note that this is always true because both compression methods use a boxcar window as a weighting function in the $uv$-space which means
that all the pre-averaged visibilities are equally weighted for both BDA and simple averaging.}
If we  \Refcom{compress} the visibilities  using a  BDWF \Refcom{$X(u,v)$} or a BDAWF $X_{\Refcom{\scalebox{0.5}{BDA}}}(u,v)$, the noise 
term still remains different per each visibility $pq\RefEq{kr}$:
\begin{equation}
\label{eq:noise:bdwf}
\sigma_{X_{pq\RefEq{kr}} }^2 = \frac{\sum X^2(\bmath{u}_{pqij}-\bmath{u}_{pq\RefEq{kr}})}
{\big [ \sum X(\bmath{u}_{pqij}-\bmath{u}_{pq\RefEq{kr}}) \big ]^2 } \, \sigma_\mathrm{s}^2,
\end{equation}
where the sums are taken over the baseline-independent \Refcom{sampling  bin indices and}  
\begin{equation}
\label{eq:noise:bdabdwf}
\sigma_{X_{pq\RefEq{kr}}, \Refcom{\scalebox{0.5}{BDA}}}^2 = \frac{\sum X_{\Refcom{\scalebox{0.5}{BDA}}}^2(\bmath{u}_{pqij}-\bmath{u}_{pq\RefEq{kr}})}
{\big [ \sum X_{\Refcom{\scalebox{0.5}{BDA}}}(\bmath{u}_{pqij}-\bmath{u}_{pq\RefEq{kr}}) \big ]^2 } \, \sigma_\mathrm{s}^2,
\end{equation}
where the sums are taken over the baseline-dependent \Refcom{sampling bins indices}. 

Eq.~(\ref{eq:noise:bdwf}) and~(\ref{eq:noise:bdabdwf}) are of critical importance on the squared error norm in each pixel of the dirty  image and so they merit detailed explanation:
\begin{enumerate}[1)]
  \item \Refcom{In $t\nu$-space}, the length of the window $X(u,v)$ \Refnew{(BDWF)} remains constant across all  baselines while the window \Refcom{resolution  varies} on different baselines: in this sense, $X(u,v)$ is baseline-dependent. Because the length of $X(u,v)$ is constant along all the baselines, the compression factor also remains constant across all the baselines, as when applying a simple averaging \Refcom{(see \citet{atemkeng2016})}.
   \item In $t\nu$-space, the window $X_{\Refcom{\scalebox{0.5}{BDA}}}(u,v)$ (BDAWF) varies in length (hence the extrat index ${\Refcom{\scalebox{0.5}{BDA}}}$) and \Refcom{resolution} across all  baselines. Because the length of $X_{\Refcom{\scalebox{0.5}{BDA}}}(u,v)$ varies along baselines, 
   the compression factor thus varies on different baselines (looking back to  Figure~\ref{fig:bda-boxcar-uvleng-directtion}).
   \item If one were to constrain the compression factor $\mathrm{CF}$ to be equal for both ``BDWF'' and ``BDAWF'', the squared error norm in each pixel of the dirty  image will change radically. This can be understood by looking at 
   steps 1) and 2): $X(u,v)$ and $X_{\Refcom{\scalebox{0.5}{BDA}}}(u,v)$ produce completely different weights for each  $(u,v)$ point. In other words, $X(u,v)\neq X_{\Refcom{\scalebox{0.5}{BDA}}}(u,v)$ for a given $(u,v)$ point.
\end{enumerate}

The visibility noise penalty induced by BDA   or BDAWF  is the relative increase in noise over simple averaging:
\begin{alignat}{2}
\Xi_{X_{\mu}} &= \frac{\sigma_{\mathrm{X_{\mu}}}}{\sigma_{\Refcom{\scalebox{0.5}{AVG}}}}.
\end{alignat}
Here, $\sigma_{\Refcom{\scalebox{0.5}{AVG}}}=\sigma_{\mathrm{s}}^2/(n_tn_{\nu})$ is the noise on the 
simple averaged visibility and $\sigma_{\mathrm{X_{\mu}}}$ is either the noise induced by BDA  or BDAWF.
The centre pixel noise penalty in the image with imaging weights $W$:
\begin{equation}
\Xi^W_{\mu} = 
\frac{\sigma_{{pix},X}^2}{\sigma_{{pix}}}  =
\frac{ (\sum_{\mu} W_{\mu}^2 \Xi_{X\mu}^2) }{ (\sum_{\mu} W_{\mu})^2}. \label{eq:noisepenalty:natural}
\end{equation}
Note that the  noise penalty properties induced by overlapping BDWFs defined in \citet{atemkeng2016} remains valid  for BDA and BDAWF.

\ATMR{Simulations confirm the theoretical noise penalty estimate discussed above. 
The  simulation consists of two datasets; the high-res  and the low-res datasets using the MeerKAT \Refnew{telescope}. 
The high-res dataset is simulated with $\sigma_s=1$ Jy thermal noise during a total period of 4 hr with 1 s integration time and 84 MHz bandwidth divided into  channels of  84 kHz. We then \Refcom{compress} the high-res using simple averaging, then BDA and BDAWFs, and save the resulting visibilities to the low-res dataset. For both \Refcom{compression} schemes, we fixed the compression factors to \CF{15}{10} and \CF{30}{20}, which then correspond to simple averaging  across \BIN{15}{0.84} and \BIN{30}{1.68} respectively. 
We use the $\Sincc$  tuned to a FoI of  $1.3^\circ$ with overlap factors of $6\times5$ of the baseline-dependent \Refcom{sampling} bins. For each case of compression, we then consider the r.m.s pixel noise as an estimator of $\sigma_{pix}$ (simple averaging) and $\sigma_{{pix},\scalebox{0.5}{X}}$ (BDA or BDAWFs). 
The analytical estimated and simulated noise penalty are compared in Table~\ref{tab:noise-comparison}.
Results confirm that
both analytical estimates and simulations agree.}

\section{Simulations and results} 
\label{results:simulation}
Having explored the mathematics and   implementation  of BDA,
we now turn to the simulation aspects. The simulations are performed with the MeerKAT and the EVN \Refnew{telescopes}. The simulated images are not calibrated and deconvolved to avoid introducing additional  effects relative to calibration and/or  deconvolution algorithms.
Two test scenarios  are considered  and both of them 
are simulated using MeqTrees~\citep{noordam2010meqtrees}:
 \begin{itemize}
  \item We consider a 1 Jy point source at various sky positions,  with no noise or other corruptions included.
We evaluate the efficiency of a BDA correlator using 
two different procedures. Firstly, we simulate the source at a fixed sky position\EDIT{,} apply BDA, BDAWFs and measure 
the \Refnew{compression} effects separately on each baseline. Secondly, we simulate the point source  at various angular distance\EDIT{s}
from the phase 
centre and apply BDA and BDAWFs, thereby  evaluating the interferometer  
cumulative decorrelation effects on all baselines. 
We measure the source 
peak amplitude in each dirty image after \Refcom{compression}. Since each dirty image corresponds to a single source, the peak gives
us the degree of smearing associated with a given \Refcom{compression} method and compression factor. 
\item \Refcom{The PB on its own could be used for source suppression, the higher the  frequency the less sources out of the FoI contaminate the image. Tests are performed when the PB is included during the simulations, BDA and BDWFs are applied
to evaluate the combined degree of suppression for sources out of the FoI.} 
\end{itemize}

\subsection{Application to MeerKAT data}
\subsubsection{Source amplitude and East-West baselines} 
\label{subsection:meerkaysupression}
\begin{table}
\begin{tabular}{lll}
\hline
{\bf Filters} & {\bf $\Xi$ theo} & {\bf $\Xi$ sim}\\
\hline\hline
\Refcom{BDA} \BIN{15}{0.84}&1.00&1.03\\
\Refcom{BDA} \BIN{30}{1.68}&1.00 &1.004\\
\hline
\WF{\Refcom{BDA}-\Sincc-}{6}{5}-1.3deg \BIN{15}{0.84} &1.19 &1.23\\
\WF{\Refcom{BDA}-\Sincc-}{6}{5}-1.3deg \BIN{30}{1.68}  &1.51&1.56\\
\hline

\end{tabular}
\caption{A comparison of image noise penalties associated with different BDA and BDAWFs, computed analytically ($\Xi$ theo)
vs. simulations ($\Xi$ sim). The analytical noise penalty for BDA is equal to 1, this is
\Refcom{straightforward} by looking at  Eq.~(\ref{noise:bda_pixel}) and~(\ref{noise:bda_pixelx1}).
}
\label{tab:noise-comparison}
\end{table}
\newcommand{\bdaWF}[3]{{#1}$#2${}$\times${}$#3$} 
\begin{figure*}
\centering
\includegraphics[width=.4\textwidth]{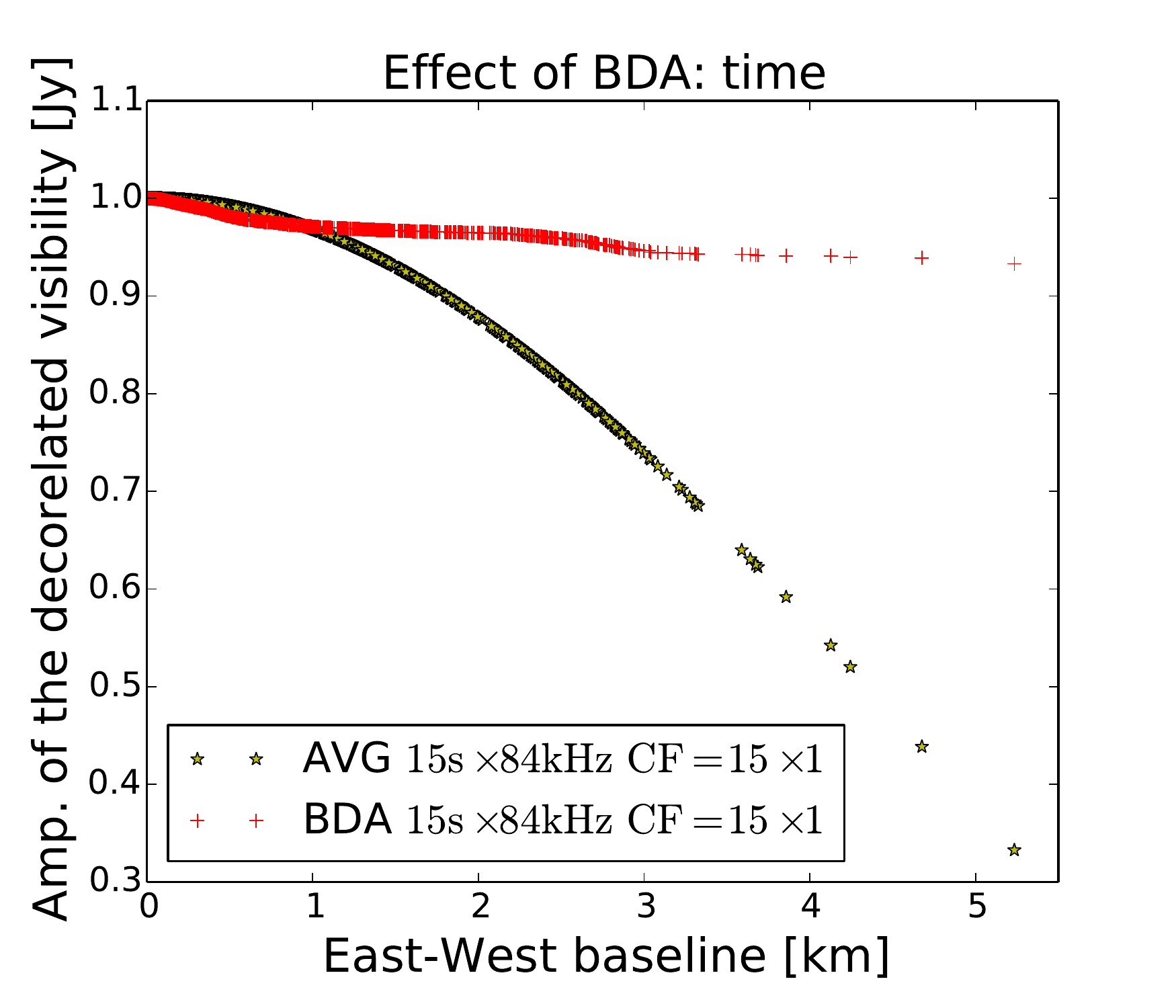}
\includegraphics[width=.4\textwidth]{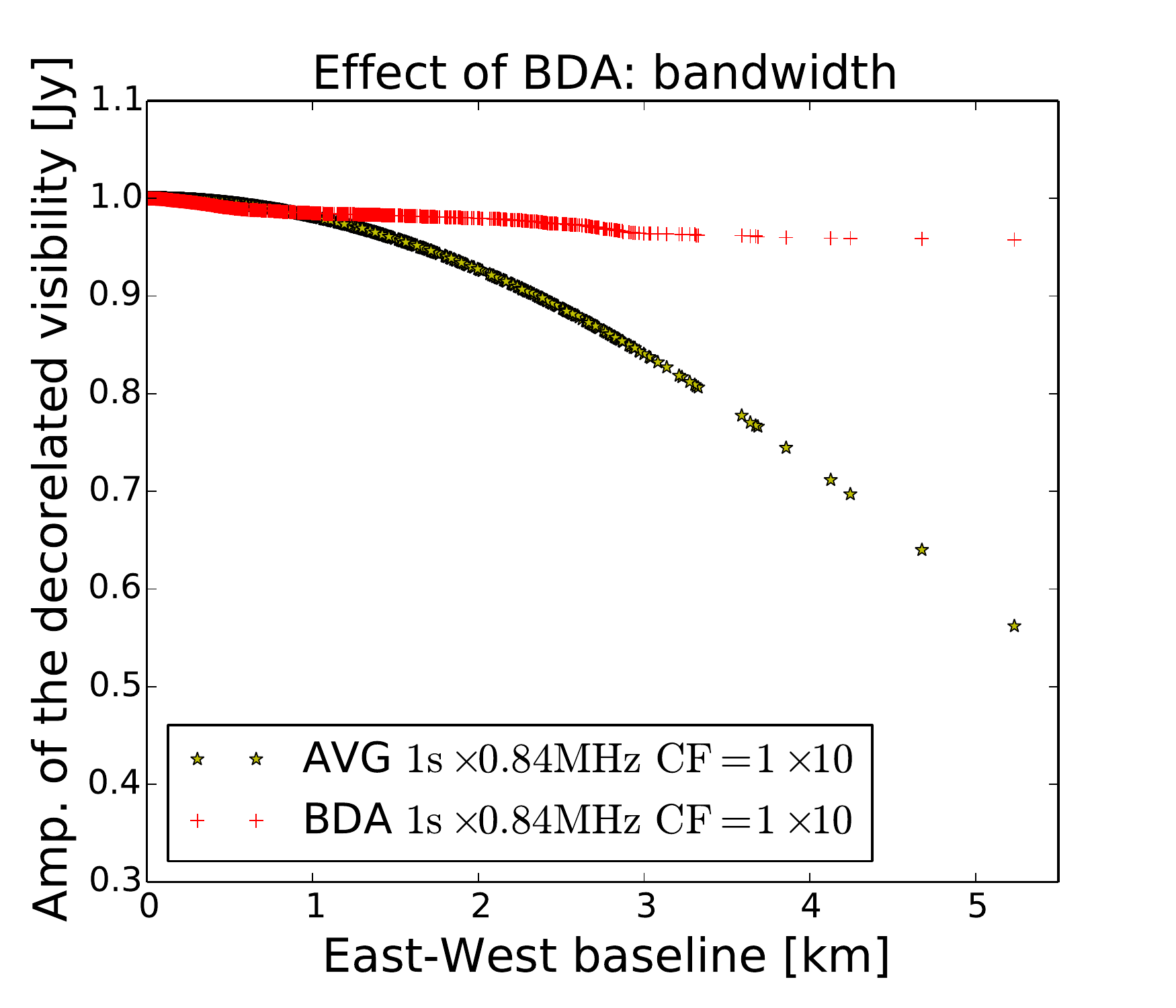}
\includegraphics[width=.4\textwidth]{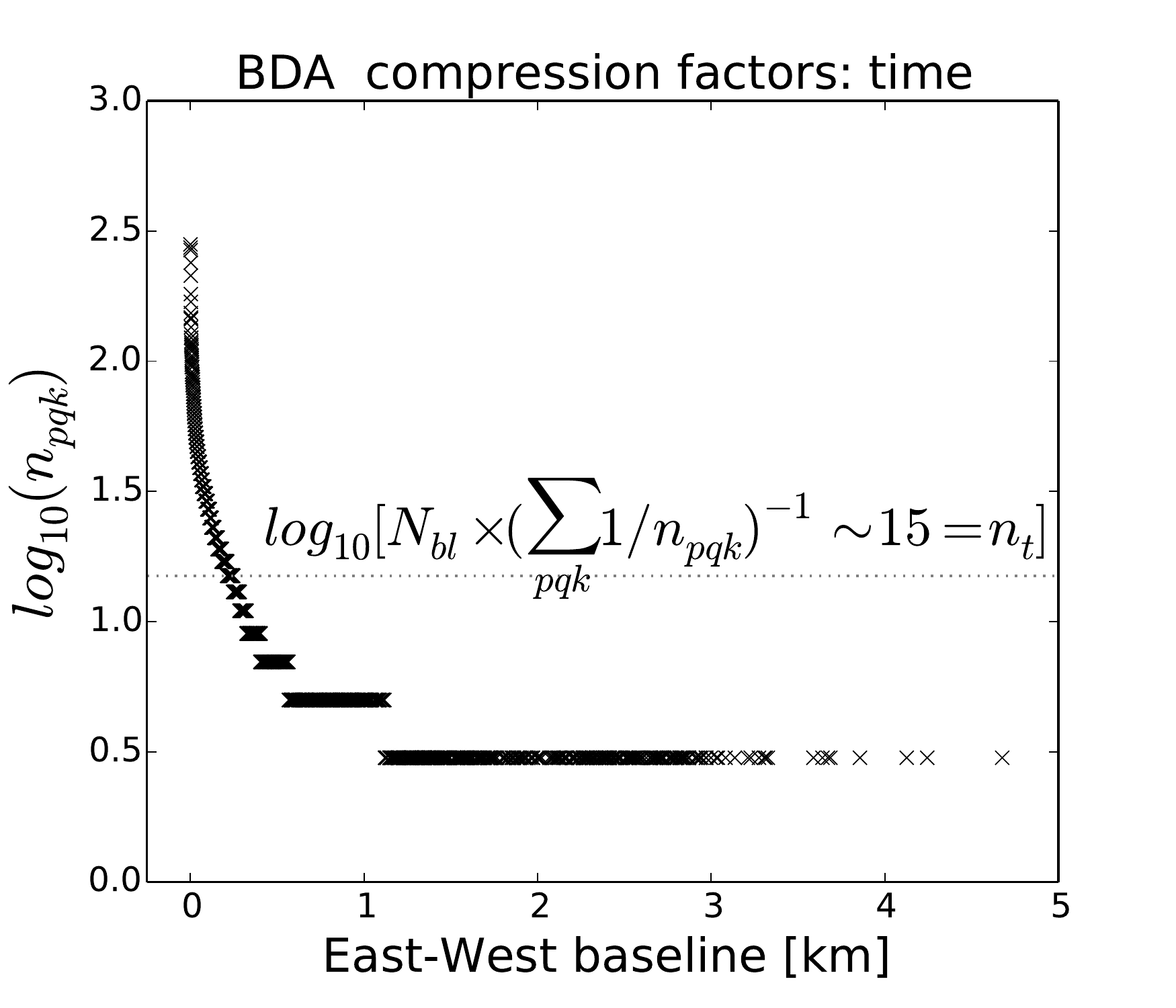}%
\includegraphics[width=.4\textwidth]{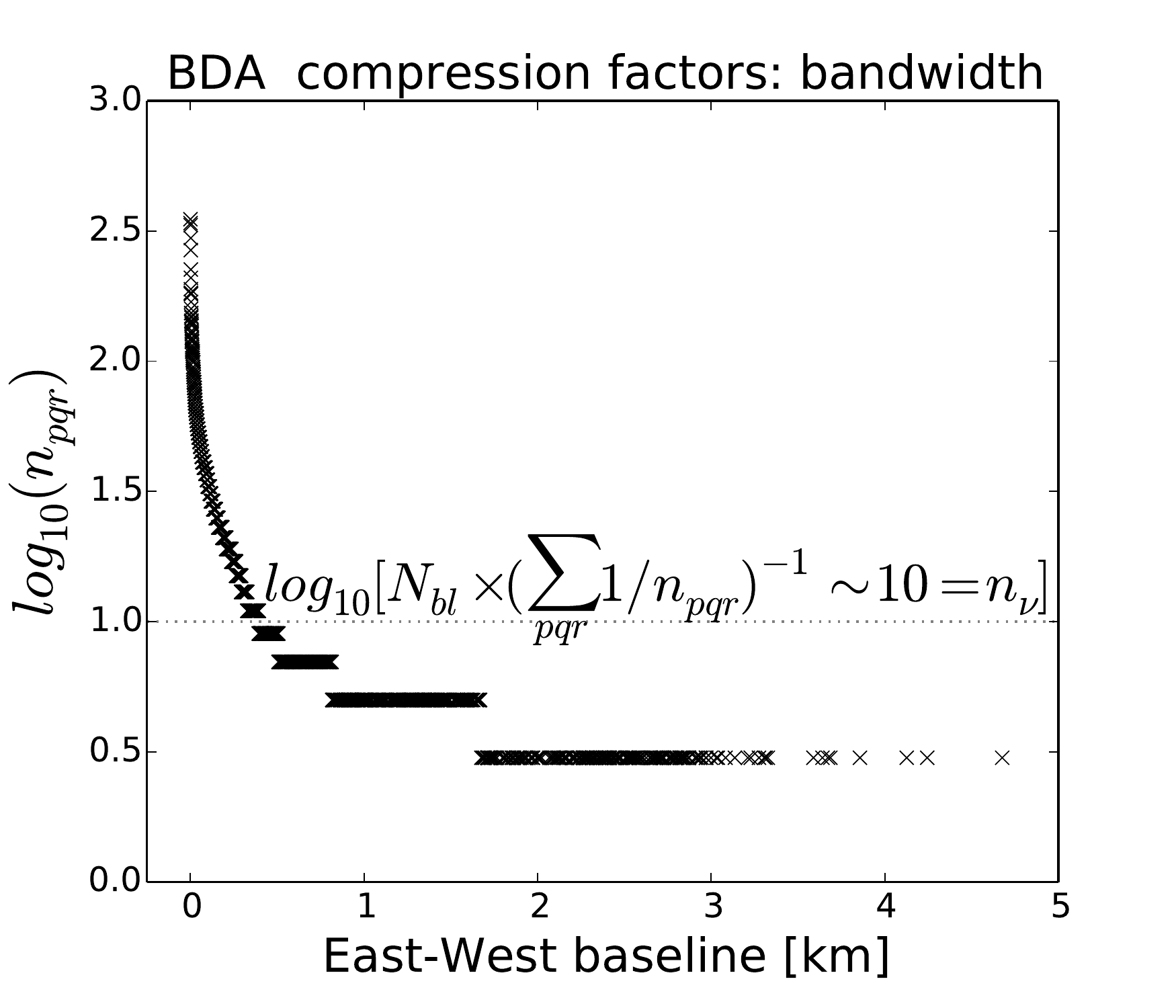}\\
\caption{(Top) Amplitude loss: the apparent intensity of a 1 Jy source at 2.25 deg as seen by MeerKAT at 1.4 GHz, 
as a function of East-West baseline components;  (left) compression carried out only in time with compression factor fixed 
to 15 time-bins; (right) compression is carried out only in frequency with compression factor fixed to 10 frequency-bins. (Bottom) Baseline-dependent  compression factors in time (left) and frequency (rigth) 
both in logarithm scale as a function of East-West baseline length. }\label{fig:srcat30arcminx}
\end{figure*}
The experiment in Figure~\ref{fig:srcat30arcmin_avg} is repeated. The simulation consists of two high-res measurement sets (MSs), each with a source at 2.25 deg relative to the observation phase centre.
Two low-res MSs are generated to receive the \Refcom{compressed} visibilities.
The results of the decorrelation when applying simple averaging and BDA  are compared in the top panel of Figure~\ref{fig:srcat30arcminx} and the BDA compression factors \Refcom{achieved with the simulation} are plotted in the bottom panel of Figure~\ref{fig:srcat30arcminx}.
 \begin{itemize}
  \item Time decorrelation and compression factors, Figure~\ref{fig:srcat30arcminx} (left): the  MS  consists of 64 frequency channels of 84 kHz width each, and
 7200 s   timeslots of 1 s integration time. The compression factor is fixed to \CF{15}{1}  \EDIT{for both} simple averaging and  BDA.
  For  BDA, the shorter baselines are compressed by a lot more than 15 and the longer baselines by a lot less, while for simple averaging this corresponds to 15 factor of compression along all the baselines.
  \item Bandwidth decorrelation and compression factors, Figure~\ref{fig:srcat30arcminx} (right):  The  MS consists of 100 timeslots of 1 s integration, and
  1000 frequency channels of 84 kHz (total bandwidth of 84 MHz).  The compression factor is fixed to \CF{1}{10} both for simple averaging and BDA.
 For  BDA, the shorter baselines are compressed by a lot more than 10 and the longer baselines by a lot less, and for simple averaging this corresponds to a compression factor of 10. 
\end{itemize}
It is  \EDIT{clearly} noticeable in the top panels of Figure~\ref{fig:srcat30arcminx} that on shorter baselines, the smearing rates of simple averaging and BDA are approximately equivalent despite the little percentage of signal lost with BDA in the region between 0.2 km and 0.8 km.  
 This can be understood by looking at the MeerKAT histogram depicted in Figure~\ref{fig:meerkat}, this is the region where one wants to compress the data as bigger as possible.
 However, for a source at 2.25 deg and at these BDA compression factors the degree of the decorrelation remains approximately equal across all the baselines. This result confirms our mathematical prediction in Eq.~(\ref{eq:bda2taper}).
 It appears from the \Refcom{simulated} time and frequency BDA compression factors  depicted in the bottom of
 Figure~\ref{fig:srcat30arcminx} that the data are compressed more in frequency than in time. This is because, for MeerKAT,  the  $uv$-track along  0.84 MHz  is smaller than the $uv$-track along 15 s. 
 We can still constrain the compression factors to be equal in both time and frequency, in principle, the shape of the 2-D $uv$-track should be square-like. To derive this, we note that the averaged bandwidth must be equal to $w_e \nu_r \Delta t$, where the constant $w_e$, is the Earth rotation velocity~\citep{thompson2001interferometry}.

\subsubsection{Source amplitude and distance from the phase centre}
\label{subsect:meerkat}

We simulate data at high time-frequency resolution of 1 s  integration during 4 hr and 84 kHz channels width 
for a total bandwidth of  84 MHz centred  at 1.4 GHz.  The sky model is a single 1 Jy point source at 
a given distance from the phase centre. Three MSs are generated to store the 
\Refcom{compressed} visibilities:
 \begin{itemize}
  \item  \textcolor{black}{Two MSs  contain the \Refcom{compressed} visibilities for  \BIN{15}{0.84} and \BIN{30}{1.68}, this result in compression factors of \CF{15}{10}  and \CF{30}{20} respectively.}
  \item A third MS to receives the \Refcom{compressed} visibilities for BDA and BDAWFs. This MS is a copy of the high-res MS where \Refnew{the flagging implementation for BDA}
    \EDIT{described} in Sect.~\ref{BDA:impl} is applied. Two compression factors \EDIT{are}  adopted for the BDA and BDAWFs:  \CF{15}{10} and \CF{30}{20}.
\end{itemize}

\begin{figure*}
\centering
\includegraphics[width=.9\textwidth]{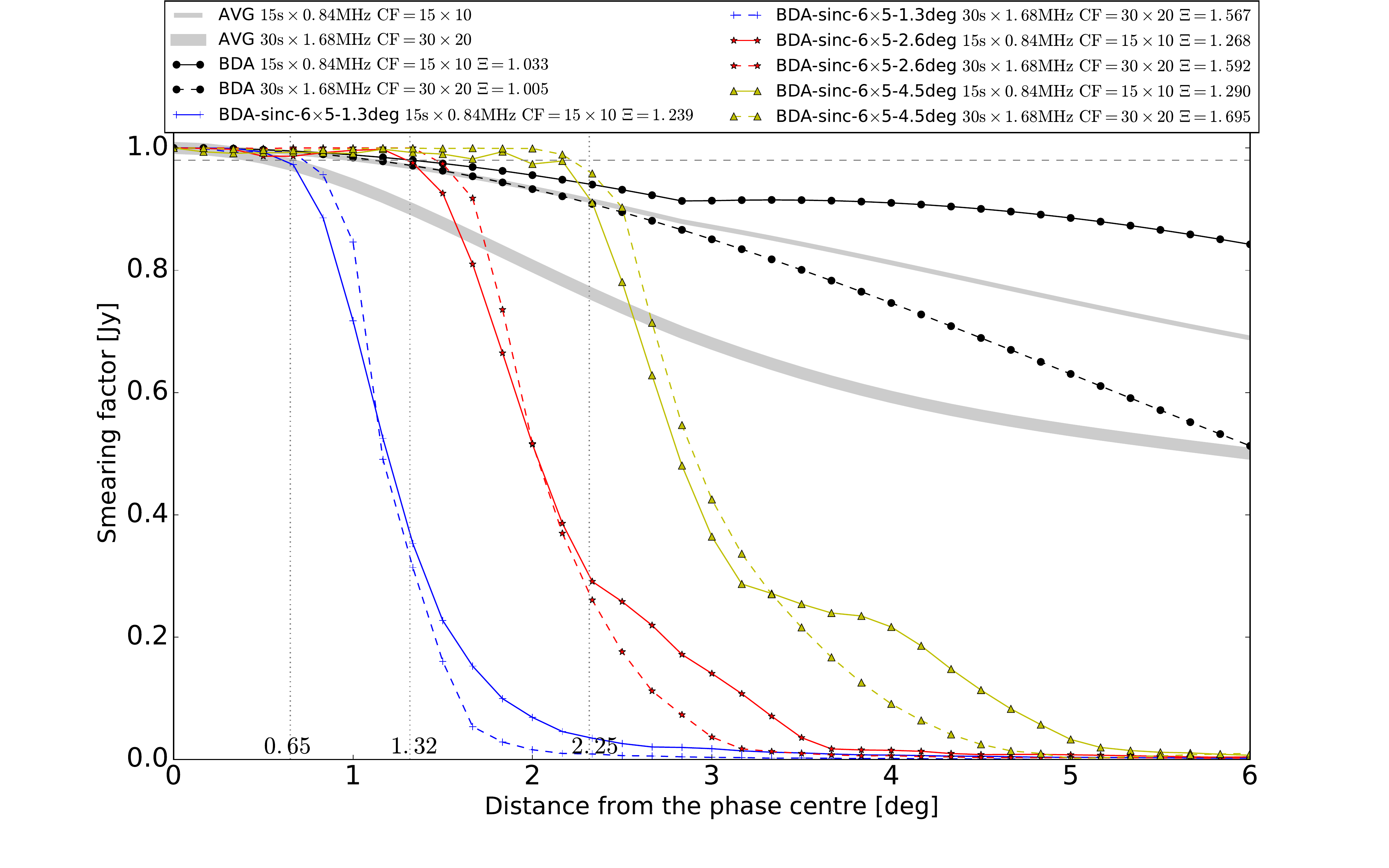}
\caption{Amplitude loss: the apparent intensity of a 1 Jy source as seen by the MeerKAT telescope at 1.4 GHz 
as a function of distance from phase centre, for simple averaging with \BIN{15}{0.84} and \BIN{30}{1.68} bins, and for BDA and BDAWFs. 
The compression factor is fixed to \CF{15}{10} and \CF{30}{20} for all the \Refcom{compression} methods.}\label{fig:bda-sn-bessel-2ge1}
\end{figure*}

Figure \ref{fig:bda-sn-bessel-2ge1} shows the performance of different \Refnew{compression schemes} and compression factors associated with their noise penalty. BDA applied to a sinc-like BDWF is considered in this test and is turned 
to three different FoI settings, as indicated by the plot: 0.65 deg,  1.32 deg and 2.25 deg. 
The results can be alternatively appreciated by regarding the performance of BDAWFs:

BDA with  \CF{15}{10} provides good results in flux recovery\EDIT{,} i.e. for $6\%$ smearing we can image up to \Refnew{4.5 deg FoI}, while  simple averaging at the same compression factor can only recover this FoI at 
$10\%$ smearing.
The BDA with compression factor \CF{30}{20} still provides better source recovery compared to simple averaging  at the same compression factor. 
We can also \EDIT{note} that at the same compression factor, the source suppression performance of BDA is worse than that of simple averaging.

\Refnew{At the different compression factors, we see that all the BDAWFs filters provide excellent performance in source recovery and far-field suppression} compared to simple averaging or BDA: smearing  across the FoI is less than  2\% (horizontal grey dashed-line), and out-of-FoI  suppression is almost two orders of magnitude higher \EDIT{than} simple averaging or BDA. Note the tapering \EDIT{behaviour} for BDAFWs at the different compression factors. As the compression factor increases, the response of BDAWFs becomes \textbf{flat}: this clearly illustrates their excellent performance. The reason for this is that, a unique sinc-like window function  is applied on all the baselines (\Refcom{recall from Figure~\ref{fig:bda-boxcar-uvleng-directtion}}). For larger  compression factors the sinc-like window function becomes more proximate to the ``sinc''\EDIT{,} which results in a more optimal ``boxcar-like'' taper in the image domain. In general, the noise penalty does depend on the compression scheme and parameters, this is the case for  BDAWF, where all the parameters i.e. compression factors, overlapping bins and FoI result in noise penalty-dependent.

\subsubsection{Relative SNRs using MeerKAT data }
\label{sect:relativeSNR}
\begin{table*}
\begin{tabular}{ |p{5cm}||p{2cm}|p{2cm}|p{2cm}|p{2cm}| }
 \hline
 \bf FILTERS&  \bf 0.65 deg&\bf 1.32 deg&\bf 2.25 deg\\
 \hline
   \Refcom{AVG} \BIN{15}{0.84}     &16.827& 16.437  &  15.672 \\
  \Refcom{BDA} \BIN{15}{0.84}   &14.767 & 14.544 &  14.072 \\
 \WF{\Refcom{BDA}-\Sincc-}{6}{5}-{1.3deg} \BIN{15}{0.84}     &64.354 &11.590&1.144  \\
  \WF{\Refcom{BDA}-\Sincc-}{6}{5}-{2.6deg} \BIN{15}{0.84}  &40.256 & 64.538& 9.554 \\
 \WF{\Refcom{BDA}-\Sincc-}{6}{5}-{4.5deg} \BIN{15}{0.84}    & 32.576&32.569&  60.249 \\
 \hline
\end{tabular}
\caption{Simulated SNR as decribed in Eq.~\Refcom{(}\ref{eq:snr}\Refcom{)}, i.e. $\mathrm{SNR}\approx S_\mathrm{smear}/(C_\mathrm{noise}+T_\mathrm{noise})$, where  $S_\mathrm{smear}$, $C_\mathrm{noise}$ and $T_\mathrm{noise}$ are defined 
as the signal of a source of interest, and the contamination signals that affect the signal of interest  and the  thermal noise respectively.
Here $T_\mathrm{noise}=\sigma_{{pix}, X}$ defined in Sect.~\ref{sect:noisepenalty}.}
\label{tab:relativeSNR}
\end{table*}
Simulations are used to separate the variables $S_\mathrm{smear}$, $C_\mathrm{noise}$ and $T_\mathrm{noise}$
in Eq.~\Refcom{(}\ref{eq:snr}\Refcom{)}. The simulated MS in Sect.~\ref{subsect:meerkat} is reused. We consider to evaluate the SNR of an image of $\sim$0.5 square degree centered at 0.65 deg, 1.32 deg and 2.25 deg. For each case, we know $S_\mathrm{smear}$  from Figure~\ref{fig:bda-sn-bessel-2ge1}.
To evaluate the contamination, and for each case, we simulate two sources: a nearby source of 1 Jy (1 deg away from each case) and a distant source of 10 Jy (20 degrees away from from each case),
and  make an image. The image will be empty, except for the contribution from these two sources. For the thermal noise, an empty sky is simulated with  1 Jy thermal noise for each of the cases listed above. 
The different compression methods are applied and their resulting SNRs are listed in Table~\ref{tab:relativeSNR}.  \ATMR{Results show that our compression technique demonstrates
better performance in SNR when compared to simple averaging. Comparatively, using BDAWFs provide the best performance in SNR, up to a factor of $\sim$4  higher than simple averaging or BDA. 
Note that in regions where the source suppression response of  BDAWFs kicks in, the SNR quickly drops, since BDAWFs are suppressing the source signal itself at this point.}

\subsubsection{ BDAWFs combined with the primary beam and source suppression}
\label{sect:beamandBDAWFs}
The additional degree of source suppression provided by BDAWFs auguments the source suppression provided by the PB, as investigated by e.g.~\citep{mort2016analysing}. Note that BDA by itself (without window functions) actually provides \lq\lq less’’ source suppression than simple averaging, at the same compression factor. 

\begin{figure*}
\centering
\includegraphics[width=1.\textwidth]{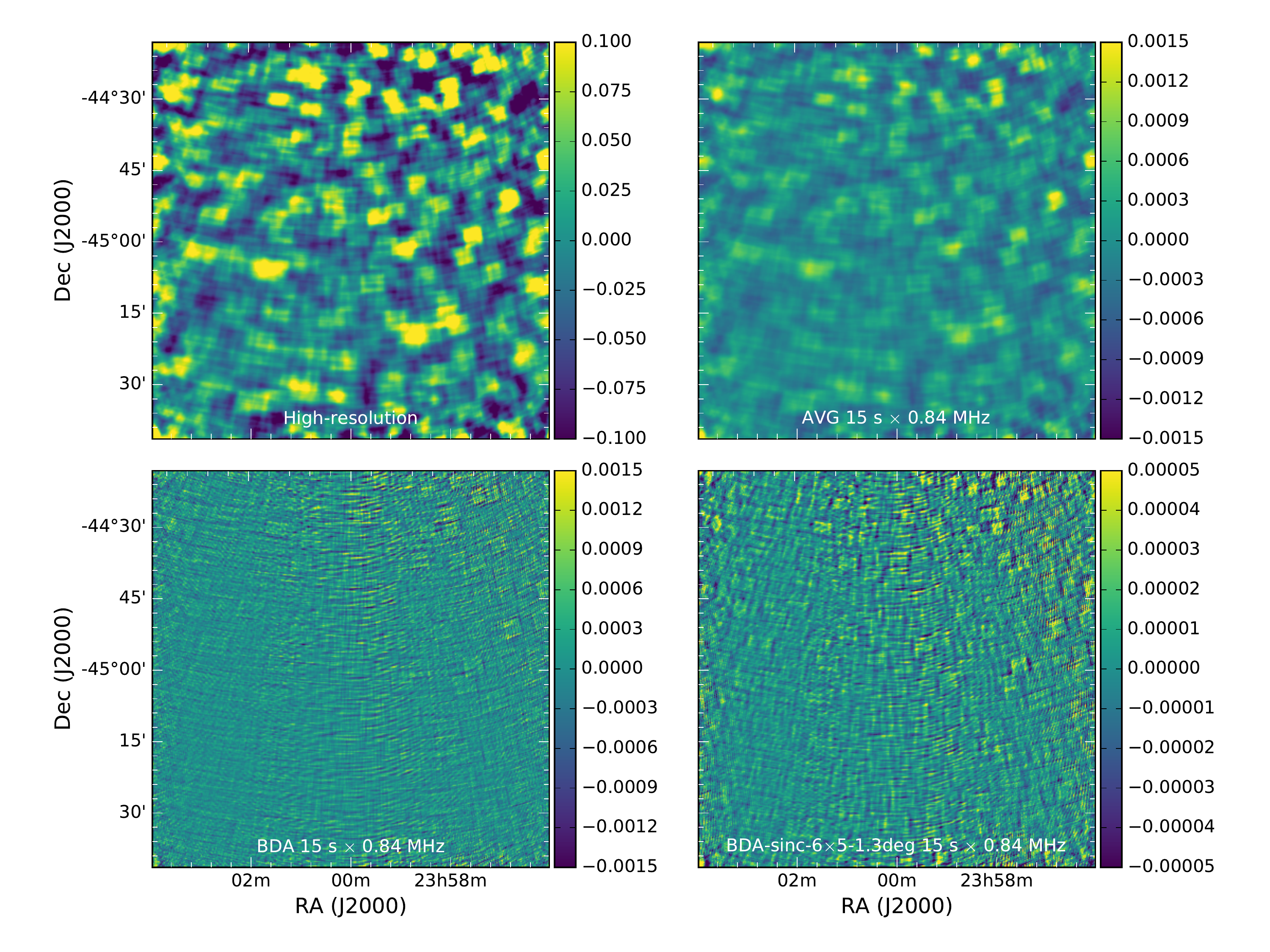}
\caption{Contamination in the FoI from a 20 Jy source located at the second null of the MeerKAT primary beam. Initially, the data is imaged without data compression been carried out (top-left panel). After  data compression is 
applied using \Refcom{AVG} \Refcom{\BIN{15}{0.84}} (top-right), \Refcom{BDA} \Refcom{\BIN{15}{0.84}} (bottom-left) and 
\Refcom{BDA}-sinc-6$\times$5-1.3deg \Refcom{\BIN{15}{0.84}} (bottom-right). \Refcom{The colourbars  of the images are in Jansky and are in different scales.} BDAWFs offer better
reduction in  source contamination compared to  \Refcom{AVG} \Refcom{\BIN{15}{0.84}} and \Refcom{BDA} \Refcom{\BIN{15}{0.84}}.}
\label{fig:srcat30arcmin_avg-x}
\end{figure*}

In this section, we investigate and compare the combined suppression factor achieved by the PB and averaging, BDA and BDAWFs.
\ATMR{
A PB model for MeerKAT at 1.4 GHz along with a \Refnew{nearby} 20 Jy source located at the second sidelobe of the PB is simulated using the MS described in  Sect.~\ref{subsect:meerkat}. 
We supposed imaging up to the FWHM of the MeerKAT  PB at 1.4 GHz (i.e. 0.65 deg away from the field centre). Three filters are considered and compared, \Refcom{AVG \BIN{15}{0.84}},  \Refcom{BDA \BIN{15}{0.84}} and  \Refcom{BDA}-sinc-6$\times$5-1.3deg \Refcom{\BIN{15}{0.84}} both having for compression factor CF=15$\times$10.
Figure~\ref{fig:srcat30arcmin_avg-x} shows dirty images of size $40\times40$ arcmin \Refcom{at different pixel scales}. These images should be empty except the contamination from the \Refcom{nearby} source.
The top-left and top-right images \Refnew{of Figure~\ref{fig:srcat30arcmin_avg-x} show} the high-res (i.e. image produced with the pre-averaged MS) and the simple averaged images  respectively.
The bottom-left and bottom-right images are produced after applying \Refcom{BDA} \Refcom{\BIN{15}{0.84}} and  \Refcom{BDA}-sinc-6$\times$5-1.3deg \Refcom{\BIN{15}{0.84}} respectively.
For both cases, the high-res image is confusion noises dominated across the FoI. The compressed images show a more confusion noise-free images.
Unlike BDA  that considers only
a flux recovery in the image domain, BDAWFs consider both flux recovery in the given FoI and source suppression out of this FoI. This is clearly seeing in Figure~\ref{fig:srcat30arcmin_avg-x} that BDA 
on its own does not remove the contamination than simple averaging but BDAWF does remarkably well. }

\subsection{BDAWFs and the EVN}
\begin{figure*}
\centering
\includegraphics[width=.9\textwidth]{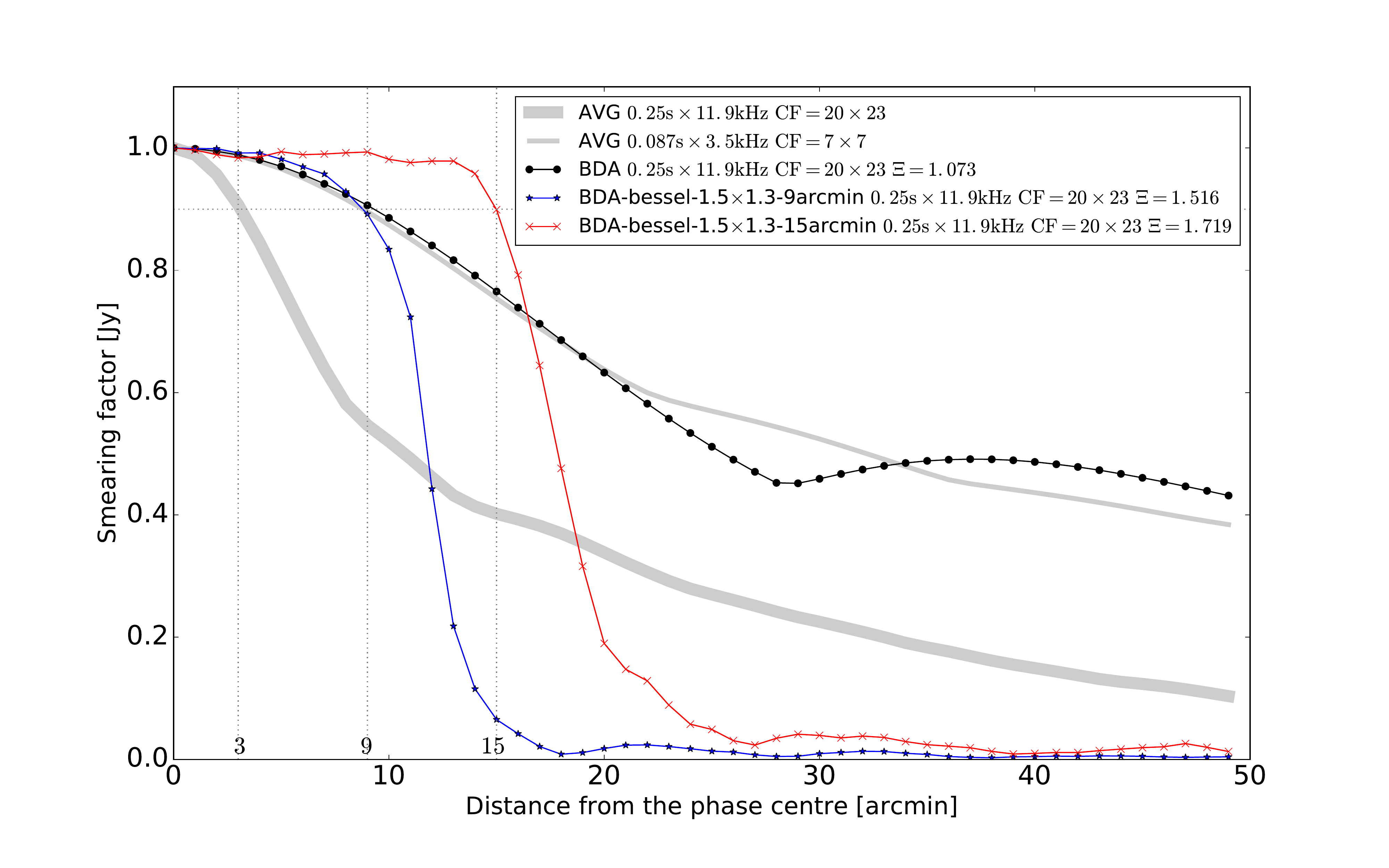}
\caption{Amplitude loss: the apparent intensity of a 1 Jy source as seen by EVN at 1.6 GHz as 
a function of distance from phase centre.  \ATMR{Results show that the data can be  compressed alot more than a factor of 9.38 using BDAWFs.}
}\label{fig:bda-sn-bessel-2ge3}
\end{figure*}

In VLBI the baselines are so long (up to $\sim$10000 km) that the FoV is always limited, and normally it is only a tiny fraction of the PB at the FWHM because of decorrelation due to time and \Refnew{bandwidth} averaging. To keep decorrelation/smearing at acceptable level one may apply wide-FoV correlation, but handling the resulting data volumes has been challenging (e.g. \citet{chi2013deep}).
Another solution is to $uv$-shift wide-field correlated data to various phase centres and apply averaging then to obtain a number of smaller FoV within the PB~\citep{morgan2011vlbi, ruiz2017faint}. This has been fully
implemented in the EVN Software Correlator (SFXC;~\citet{keipema2015sfxc}). Multi-phase centre correlation makes milliarcsecond-resolution imaging of a-priori known sources spread
over a wide-\Refnew{FoV} possible, this has now been applied
routinely at the EVN.
But some applications (e.g. transient search within the full PB in
VLBI data, or to
build up a wide-FoV EVN archive) would require storing the raw data
from all telescopes,
however, this results in very large volumes unless there are
alternative approaches.
We investigate the possibility of using BDA and BDAWFs in VLBI to
preserve a significant
fraction of the PB while significantly reducing the data volume. We repeated the
simulation scenarios described in Sect.~\ref{subsect:meerkat} using the full EVN (i.e.
Badary, Effelsberg, Hartebeesthoek, Jodrell Bank, Medicina, Noto,
Onsala, Shanghai, Svetloe, Torun, Westerbork, Zelenchukskaya) at 1.6
GHz. The results are given in Figure~\ref{fig:bda-sn-bessel-2ge3}. It can be seen
that for a certain compression rate with simple averaging that would
result in a FoI of
6 arcmin, an equivalent compression rate using BDA or BDAWFs would
result in a FoI
of 18 arcmin. We also note that, if one aims at imaging a FoI of
18 arcmin with
simple averaging, then this is possible with BDA reducing data by a factor of 9.38,
and the factor can be even higher with BDAWFs. While these initial
tests are very
promising, in VLBI there is a significant trade-off in sensitivity and
resolution,
therefore the best approach should be investigated in detail
independently for each
science application.

\section{Conclusion and perspectives}
\label{conclusion}
As discussed above compression of  visibilities by simple averaging  shows that
decorrelation/smearing is more significant on longer baselines than shorter ones and that decorrelation can only be avoided if the correlator performs the averaging procedure over \Refcom{smaller}  bins, which however results in high
data rates. 
We now make predictions pertaining to \Refcom{sample the visibilities regularly across all the baselines in the entire $uv$-space and apply BDA and BDAWFs.} Intuitively, in the time-frequency space (or the correlator domain) this corresponds to averaging 
within  sufficiently large \Refcom{sampling bin}  for shorter baselines, while  the longer baselines are 
averaged within shorter \Refcom{sampling bin}. 
The  question is then whether such averaging technique 
will not only decrease smearing within the observation \EDIT{FoI}, but, also reduce the data size. 
The second \EDIT{question} pertains to  calibration issues for this method given that calibration
is a complex visibilities correction process.
BDA  could introduce complexity further down the line: it could, for example, mean that a dynamic calibration solution interval would become necessary.
This implies that the calibration solution interval will 
change differently with baselines and  each of the frequency and/or time intervals.


We have established that BDA  by itself
can only achieve data compression but not FoI shaping: BDA does
decrease smearing over the  FoI, while on the other hand, sources out-of-FoI are not suppressed compared to simple averaging. 
We have found that \Refnew{BDAWFs} result \EDIT{in}  excellent tapering behaviour,
which can decrease smearing to about $2\%$ or less over a selected  FoI, with  out-of-FoI source suppression
almost two orders of magnitude higher \EDIT{than} simple averaging, while the data \EDIT{are} compressed at the same rate.


We should note that like simple averaging, BDA and BDAWFs also distort the point spread function (PSF), \Refnew{which becomes position dependent} and reacts differently compared to simple averaging. However, for an efficient use of BDA \Refnew{and} BDAWFs, one requires to \EDIT{predict} this PSF at different sky positions during deconvolution. \Refnew{There exist\Refcom{s} a faceting imaging framework that accounts for this PSF variation during deconvolution when applying BDA (see DDFacet~\citep{ctasseDDFacet}). DDFacet uses the brute-force approach to compute the PSF at the centre of each facet, and this PSF is used to deconvolve the facet. However, a brute-force computing load is tolerable for small size facets. For large facets and for any non-faceting deconvolution algorithm an approximation based method to derive all these PSFs must be implemented with the aims to reduce computing cost~\citep[in prep.]{atem2}.}

This paper opens up several possibilities for future work. Firstly,
designing an optimally matched filter for a BDAWF is an interesting avenue of further
research. In practical situations, the IPR of a sinc-like lowpass filter
is far from ideal in the sense that a sinc-like filter is band-limited (zero outside some
intervals) and sampled. Filter design theory for lowpass filters could, therefore, be used
to explore an ideal IPR, by using an approximation to define the ideal
\Refnew{filter} coefficients and parameters, such as the passband, the transition band and the stopband.
The second avenue involves evaluating the degree of source suppression as a function of
array layout and \Refnew{BDAWFs} parameters, i.e. the passband, transition band, stopband and
the size of the filter.
The third avenue of exploration consists of investigating and exploring calibration with
BDA and BDAWFs. Currently, BDA and BDAWFs can
only be used post-calibration. Exploring the calibration parameters for BDA and BDAWFs
could open a new research avenue in radio interferometry, in view of the effective use of
BDA and BDAWFs. 
Another possible work on BDA will be to explore  a possible new \Refnew{storage scheme} to take full advantage of the compression capabilities of BDA. In this work, we have considered and used only  data structures that a MS and other software packages we used can support. The \Refcom{MS has} a lot of flagging entries that still reside in memory.

Finally, this document was restricted to simulations. The next step will be to implement each
of the techniques presented in this work in practical research scenarios, e.g. applying the
filters to real interferometric data.

\section*{Acknowledgements}
This work is based upon research supported by the South African Research Chairs Initiative of the Department of 
Science and Technology and National Research Foundation. The European VLBI Network is a joint facility of independent European,
African, Asian, and North American radio astronomy institutes. M. Atemkeng is grateful to   Tammo Jan
Djimeka
for valuable discussions on BDA and its applicability to LOFAR real data during his  visit at ASTRON. The visit to ASTRON was
made possible by the FP7 MIDPREP program.
We
thank our colleagues Kshitij Thorat, Modhurita Mitra, Etienne Bonnassieux, Diana G. Klutse and Sphesihle Makhathini  for their insights  and comments on early drafts of this paper.
We would also like to thank  Khan Asad for making available the MeerKAT primary beam model used in Sect.~\ref{sect:beamandBDAWFs}.
\Refnew{The authors would like to thank the  reviewer for his valuable comments and suggestions that strongly improved the quality of the paper.}
\bibliographystyle{mn2e}
\bibliography{m_paper}

 \appendix
\section*{Appendix A: Mathematical details for the noise variance}
 \label{appendixA}
 Assuming an uncorrelated noise across the BDA averaged visibilities, the variance in each pixel in the uncleaned map for any weighting scheme $W$ is derived as:
\begin{equation}
\label{eq:noise:imageapend}
\sigma_{pix,\Refcom{\scalebox{0.5}{BDA}}}^2 = \frac{ \sum_{pq\RefEq{kr}} W_{pq\RefEq{kr}}^2 \sigma_{pq\RefEq{kr},\Refcom{\Refcom{\scalebox{0.5}{BDA}}}}^2 }{ (\sum_{pq\RefEq{kr}} W_{pq\RefEq{kr}})^2 },
\end{equation}
Setting $W\equiv1$ (natural weighting), we have:
\begin{alignat}{2}
\label{eq:noise:imageApend}
\sigma_{pix,\Refcom{\Refcom{\scalebox{0.5}{BDA}}}}^2 = \frac{ \sum_{pq\RefEq{kr}}\sigma_{pq\RefEq{kr},\Refcom{\footnotesize{\Refcom{\scalebox{0.5}{BDA}}}}}^2 }{(N_\mathrm{vis}^\mathrm{lores})^2 },
\end{alignat}
where $N_\mathrm{vis}^\mathrm{lores}$ is the total number of visibilities interring the $uv$-space after BDA. Recall that
$\sigma_{pq\RefEq{kr},\Refcom{\Refcom{\scalebox{0.5}{BDA}}}}^2 = \sigma_\mathrm{s}^2/n_{pq\RefEq{kr}}$ see Eq.~\Refcom{(}\ref{noise:avgbin}\Refcom{)}. Eq.~\Refcom{(}\ref{eq:noise:imageApend}\Refcom{)} \Refnew{leads} to:
\begin{equation}
\sigma_{pix,\Refcom{\Refcom{\scalebox{0.5}{BDA}}}}^2 = \Bigg(\frac{\sigma_\mathrm{s}}{N_\mathrm{vis}^\mathrm{lores}}\Bigg)^2\sum_{pq\RefEq{kr}} \frac{1}{n_{pq\RefEq{kr}}}.\label{noise:bda_pixelAppend}
\end{equation}
We defined in Eq.~\Refcom{(}\ref{compresionbdafactorx}\Refcom{)} the compression factor 
as $\mathrm{CF}_\mathrm{} =N_\mathrm{vis}^\mathrm{hires}/N_\mathrm{vis}^\mathrm{lores}$, thus $N_\mathrm{vis}^\mathrm{lores}  =N_\mathrm{vis}^\mathrm{hires}/\mathrm{CF}_\mathrm{}$ replacing the latter in Eq.~\Refcom{(}\ref{noise:bda_pixelAppend}\Refcom{)}, we then have:
\begin{equation}
\sigma_{pix,\Refcom{\Refcom{\scalebox{0.5}{BDA}}}}^2 = \Bigg(\frac{\mathrm{CF}\sigma_\mathrm{s}}{N_\mathrm{vis}^\mathrm{hires}}\Bigg)^2\sum_{pq\RefEq{kr}} \frac{1}{n_{pq\RefEq{kr}}}\label{noise:bda_pixelbdaAppend},
\end{equation}
which is the result in Eq.~\Refcom{(}\ref{noise:bda_pixel}\Refcom{)}.
In the case of simple averaging, where the time-frequency compression factor remains constant across all the interferometer baselines (i.e $n_{pq\RefEq{kr}}=n_tn_{\nu}=\mathrm{CF}$), the sum in Eq.~\Refcom{(}\ref{noise:bda_pixelbdaAppend}\Refcom{)} will now yield to:
\begin{alignat}{2}
\sum_{pq\RefEq{kr}} \frac{1}{n_{pq\RefEq{kr}}}&=\frac{1}{n_t n_{\nu}}N_\mathrm{vis}^\mathrm{lores}\label{averagingSum}.
\end{alignat}
If one replace Eq.~\Refcom{(}\ref{averagingSum}\Refcom{)} in Eq.~\Refcom{(}\ref{noise:bda_pixelbdaAppend}\Refcom{)} then we have:
\begin{equation}
\sigma_{pix,\Refcom{\scalebox{0.5}{AVG}}}^2 = \frac{\mathrm{CF}^2}{N_\mathrm{vis}^\mathrm{hires}} \frac{N_\mathrm{vis}^\mathrm{lores}}{N_\mathrm{vis}^\mathrm{hires}}\frac{1}{n_t n_{\nu}}\sigma_\mathrm{s}^2,
\end{equation}
knowing that $\mathrm{CF}=n_t n_{\nu}=N_\mathrm{vis}^\mathrm{hires}/N_\mathrm{vis}^\mathrm{lores}$, after simplifications we then arrived at:
\begin{alignat}{2}
\sigma_{pix,\Refcom{\scalebox{0.5}{AVG}}}^{2} &= \frac{\mathrm{CF}}{N_\mathrm{vis}^\mathrm{hires}n_tn_{\nu}}\sigma_{\mathrm{s}}^2\\
	      &=\frac{1}{N_\mathrm{vis}^\mathrm{hires}}\sigma_{\mathrm{s}}^2 
	      =\frac{1}{N_\mathrm{vis}^\mathrm{lores}n_tn_{\nu}}\sigma_{\mathrm{s}}^2,\label{noise:bda_pixelx}
\end{alignat}
which is the result presented in Eq.~\Refcom{(}\ref{noise:bda_pixelx1}\Refcom{)}, where $N_\mathrm{vis}^\mathrm{lores}$ is simply the number of visibilities in the simple averaged data, i.e. $N_\mathrm{vis}^\mathrm{lores}=N_\mathrm{vis}^\mathrm{\Refcom{\scalebox{0.5}{AVG}}}$.
\\

\bsp
\label{lastpage}
\end{document}